\documentclass[sigconf]{acmart}
\makeatletter
\renewcommand\@formatdoi[1]{\ignorespaces}
\makeatother
\usepackage{flushend}

\renewcommand\footnotetextcopyrightpermission[1]{} 
\usepackage{booktabs} 

\usepackage{hyperref}
\usepackage{booktabs,multirow}
\usepackage{gensymb}
\usepackage{textcomp}

\usepackage{mathtools}

\usepackage{color}
\definecolor{gray}{rgb}{0.5,0.5,0.5}
\definecolor{darkgreen}{rgb}{0,0.6,0}
\definecolor{commentcolor}{rgb}{0,0.6,0}
\usepackage{caption}
\usepackage{subcaption}
\usepackage{xargs}
\usepackage{xcolor}
\usepackage[colorinlistoftodos,prependcaption,textsize=tiny]{todonotes}
\newcommandx{\improvement}[2][1=]{\todo[disable,linecolor=red,backgroundcolor=red!25,bordercolor=red,#1]{#2}}
\newcommandx{\change}[2][1=]{\todo[linecolor=blue,backgroundcolor=blue!25,bordercolor=blue,#1]{#2}}
\newcommandx{\info}[2][1=]{\todo[linecolor=OliveGreen,backgroundcolor=OliveGreen!25,bordercolor=OliveGreen,#1]{#2}}
\newcommandx{\thiswillnotshow}[2][1=]{\todo[disable,#1]{#2}}

\usepackage{listings}
\newcommand{\listing}[1]{listing~\ref{#1}}
\newcommand{\Listing}[1]{Listing~\ref{#1}}

\newcounter{attcntr}
\newcommand{\newattcntr}[1]{\refstepcounter{attcntr}\label{#1}\theattcntr}

\usepackage{url}
\usepackage{balance}
\usepackage{graphicx}

\usepackage{epsfig,endnotes}

\setcopyright{none}

\acmDOI{}

\acmISBN{}

\acmConference[]{}{}{} 
\acmYear{}
\copyrightyear{}
 \renewcommand\footnotetextcopyrightpermission[1]{}

\acmArticle{}
\acmPrice{}

\begin{document}
\title[The Heisenberg Defense: Proactively Defending SGX Enclaves]{The Heisenberg Defense: Proactively Defending SGX Enclaves against 
Page-Table-Based Side-Channel Attacks}

\author{Raoul Strackx}
\affiliation{%
  \institution{imec-DistriNet, KU Leuven}
  \streetaddress{Celestijnenlaan 200A}
  \city{Leuven} 
  \state{Belgium} 
  \postcode{3001}
}
\email{raoul.strackx@cs.kuleuven.be}

\author{Frank Piessens}
\affiliation{%
  \institution{imec-DistriNet, KU Leuven}
  \streetaddress{Celestijnenlaan 200A}
  \city{Leuven} 
  \state{Belgium} 
  \postcode{3001}
}
\email{frank.piessens@cs.kuleuven.be}
%

\begin{abstract}
Protected-module architectures (PMAs) have been proposed
to provide strong isolation guarantees, even on top of a compromised system.
Unfortunately, Intel SGX -- the only publicly available high-end PMA -- 
has been shown to only provide limited isolation. 
An attacker controlling the untrusted page tables, can learn enclave secrets
by observing its page access patterns.

Fortifying existing protected-module architectures in a real-world setting against side-channel 
attacks is an extremely difficult task as system software (hypervisor, 
operating system, \ldots) needs to remain in full control over the underlying hardware.  
Most state-of-the-art solutions propose a reactive defense that
monitors for signs of an attack. Such approaches unfortunately cannot
detect the most novel attacks, suffer from false-positives, and 
place an extraordinary heavy burden on enclave-developers when an attack is detected.

We present Heisenberg, a {\em proactive} defense that provides
complete protection against
page table based side channels.
We guarantee that any attack will either be prevented or detected automatically 
before {\em any} sensitive information leaks. 
Consequently, Heisenberg can always securely resume enclave execution -- 
even when the attacker is still present in the system.

We present two implementations. Heisenberg-HW relies on very limited hardware features to
defend against page-table-based attacks. We use the x86/SGX platform as an example, but 
the same approach can be applied when protected-module architectures are ported to 
different platforms as well. 

Heisenberg-SW avoids these hardware modifications and can readily be applied. 
Unfortunately, it's reliance on Intel Transactional 
Synchronization Extensions (TSX) may lead to significant performance overhead under real-life conditions.
\end{abstract}

%
%
%
%
%

\maketitle

\section{Introduction}
Hardware security solutions such as smart cards and one-time password (OTP) generators are generally
considered more secure than their software-based implementations. Being physically separate 
devices with only a very limited physical interface without a network connection, 
they are virtually immune to malware. 
Unfortunately, they also have significant disadvantages: they are much more expensive than software,
are difficult to replace in case of loss or theft, and most devices are tight to only one specific 
service leading to a proliferation of devices a user needs to carry along.

Protected-module architectures (PMAs)~\cite{mccune2008Flicker,sahita2009SEZ,sahita2009PMAPS,
mccune2010TrustVisor,elDefrawy2012smart,strackx2012fides,champagne2010bastion,azab2011sice,
singaravelu2006ReducingTCB,noorman2013sancus,evtyushkin2014isoX,costan2015sanctum,
koeberl2014trustlite,shinde2015podarch,chen2008overshadow,
brasser2015tytan} have been proposed to bridge this gap. 
Such architectures enable small, mutually distrusting, ``protected modules'' to be isolated from 
large -- potentially compromised -- applications, operating systems 
and even more privileged layers. 
Such modules can only be 
called through specific entry points. Any other access attempt to these memory regions
is prevented. Services provided by legacy operating systems and hypervisors are still 
used, but an attacker at this level should not be able to affect the security of 
protected modules.

Early designs of PMAs leveraged their small trusted computing base (TCB) to
provide strong protection of security-sensitive ``Pieces of Application Logic'' 
(PALs)~\cite{mccune2008Flicker,mccune2009SafePassage}, 
or enable a software version of existing hardware-based security 
solutions~\cite{hoekstra2013sgxUseCases,brekalo2016passwords,krawiecka2016passwords,
himanshu2016fTPM}. 
Security of these modules is paramount. Hence, these modules tend to have a 
small codebase to make formal verification a reasonable option.

With the arrival of Intel SGX -- an Intel-implemented PMA -- a second research track 
emerged that aims to protect users from malicious cloud 
providers~\cite{baumann2014haven}.
By relying on SGX' guarantee that protected ``enclaves'' 
never reside in plaintext outside the CPU package, even an attacker with physical access
cannot directly extract sensitive data.
TCB reduction and formally verifiable security is less importance than 
ease of use. Prototypes have been developed that protect complete 
applications~\cite{baumann2014haven,graphene} or 
containers~\cite{arnautov2016scone}. 

Protected-module architectures rely on a delicate balance of responsibility between 
system software (e.g, kernel, hypervisor, \ldots) and the modules
that execute on the device. On one hand the PMA needs to ensure that 
modules execute in complete isolation and can provide strong security guarantees. 
But on the other, it is paramount that privileged system software remains in full control 
over the platform. 
A malicious or buggy protected module, for example, must not be able to make the system 
non-responsive by entering an infinite loop. Hence Intel 
SGX~\cite{hoekstra2013sgxUseCases,mckeen2013sgxInstructions,
anati2013sgxAttestationAndSealing}
ensures that its protected ``enclaves'' can be interrupted at any moment in time.

A similar but much harder problem to solve arises in the memory allocation mechanism of 
SGX enclaves. To prevent that badly behaving enclaves negatively affect the overall system, 
system software should remain in control over SGX reserved memory allocation. 
SGX fulfills this requirement
by enabling privileged software to swap enclave pages in and out of memory. Confidentiality, 
integrity and freshness of these pages is guaranteed directly by hardware.

Unfortunately, this enables powerful
side-channel attacks~\cite{xu2015controlledChannel,vanbulck2017pfless}. 
An executing enclave that attempts to access a page that is no longer 
in memory, will cause a page fault. It is up to the kernel to swap the required page back in, 
and resume the enclave. In order to do so, the address of the memory page needs to be 
provided. By carefully swapping specific pages out of enclave memory, 
sensitive information about the inner state of the enclave can 
be learned.

Existing security measures are split in two lines of research. One line focuses on the 
detection of an ongoing attack. 
Shih~et~al.~\cite{shih2017tsgx} present
in their NDSS'17 paper a solution to intercept page faults. Chen et~al.~\cite{chen2017dejavu} 
detect unexpected enclave exits through a trusted timing thread. While stopping the 
attacks described by Xu~et~al., enclave accesses are disclosed in other ways as well 
(e.g., accessed and dirty bits set in the untrusted page) 
tables~\cite{costan2015sanctum,costan2016sgxExplained,vanbulck2017pfless,wang2017leakyCauldron}. 

Another line of research relies on heavy code transformations. 
Shinde~et~al.~\cite{shinde2016pidgeonHoles} proposes to add dummy instructions and 
a scratch pad to ensure that all execution paths 
lead to identical memory access patterns. Programmer support is required to reduce overhead and 
non-balanced execution
paths are not supported. Similarly, advice by Intel~\cite{intel2017SGXDevGuide} 
to align code and data to a single
page, or by the cryptographic community to avoid sensitive data-dependent memory-accesses
are very hard to enforce in practice. Even in widely-used, hardened cryptographic libraries, 
such side-channels are present.~\cite{shinde2016pidgeonHoles}


We propose Heisenberg, a proactive defense providing {\em complete protection} against page-table 
based side channels for enclaved security-sensitive pieces of application logic.  
By preloading enclave page addresses in 
the TLB upon enclave (re-)entry, an attacker is no longer able to observe 
enclave page accesses. During an attack either the adversary can no longer 
detect which pages were accessed,
or enclaves exit automatically and immediately.\footnote{Given the similarity to
quantum physics where the location and momentum of particles cannot be determined
at the same time, we call this the Heisenberg defense.} 
As no information about 
an enclave's memory accesses ever flows to an attacker, enclaves can always be
 restarted securely, even when an attacker may subsequently continue her attack.
This avoids the significant shortcomings of state-of-the-art reactive approaches 
(see Section~\ref{sec:securityGuarantees}).
 

%


We make the following contributions:

\begin{itemize}
 \item We propose a proactive defense mechanism for
protected modules that require {\em complete} protection against page-table-based side-channel attacks.
 \item Our Heisenberg-HW prototype enables a framework to defend against many side-channel
 attacks (e.g., cache attacks). It relies only on small hardware modifications that are limited to the 
 implementation of the protected-module architecture and has a very limited performance overhead. 
 \item For readily available platforms, we present Heisenberg-SW, 
 a prototype relying on Intel's Transactional Synchronization Extensions (TSX).
 \item We investigate the use of Intel TSX as a defense mechanism to hook SGX enclave resumes and 
 show that the transaction abort rate can increase up to two orders of magnitude when the 
 system comes under heavy load, even when transactions are running on a reserved physical core and 
 interrupts are offloaded to a different core. This impacts {\em all} defense 
 mechanisms~\cite{shih2017tsgx} -- including Heisenberg-SW -- that take this approach. Our proposed hardware 
 modifications avoid such limitations.
%
%
\end{itemize}

The remainder of the paper is structured as follows. Section~\ref{sec:problemStatement} describes how 
page-table based attacks operate and state which security and operational guarantees
solutions need to provide if they are ever to be applied outside the lab. 
Our Heisenberg defense and its two variations are presented in 
Section~\ref{sec:heisenberg}. Sections~\ref{sec:implementation} and 
\ref{sec:evaluation} discuss and evaluate our two implementations. 
We discuss the impact of enclave size on 
page-table-based side channels in general and our defense in detail in 
Section~\ref{sec:discussion:enclaveSizes}. 
Finally 
we discuss related work and conclude in Sections~\ref{sec:relatedWork} and \ref{sec:conclusion}.
We refer readers with limited knowledge of Intel SGX to appendix~\ref{sec:background}.


\section{Problem Statement}
\label{sec:problemStatement}


%

SGX leaves system software in full control over virtual address space translations 
and allocation of SGX reserved memory.
Unfortunately, Xu~et~al.~\cite{xu2015controlledChannel} showed that this enables powerful 
side channels against enclaves. 
In Section~\ref{sec:attacks} we discuss their attack in a very basic form and elaborate 
on closely related attacks using alternative information leaks. Next, we more precisely define 
the attack model we need to defend against. Finally in Sections~\ref{sec:securityGuarantees} 
and \ref{sec:operationalGuarantees} 
we discuss Heisenberg's security properties and how any practical solution 
to the page-table-based side channel problem needs to ensure that
system software is able to manage the platform effectively.

%

\subsection{Page-Table-Based Side-Channels}
\label{sec:attacks}
\begin{figure}
    \centering
    \includegraphics[width=0.45\textwidth]{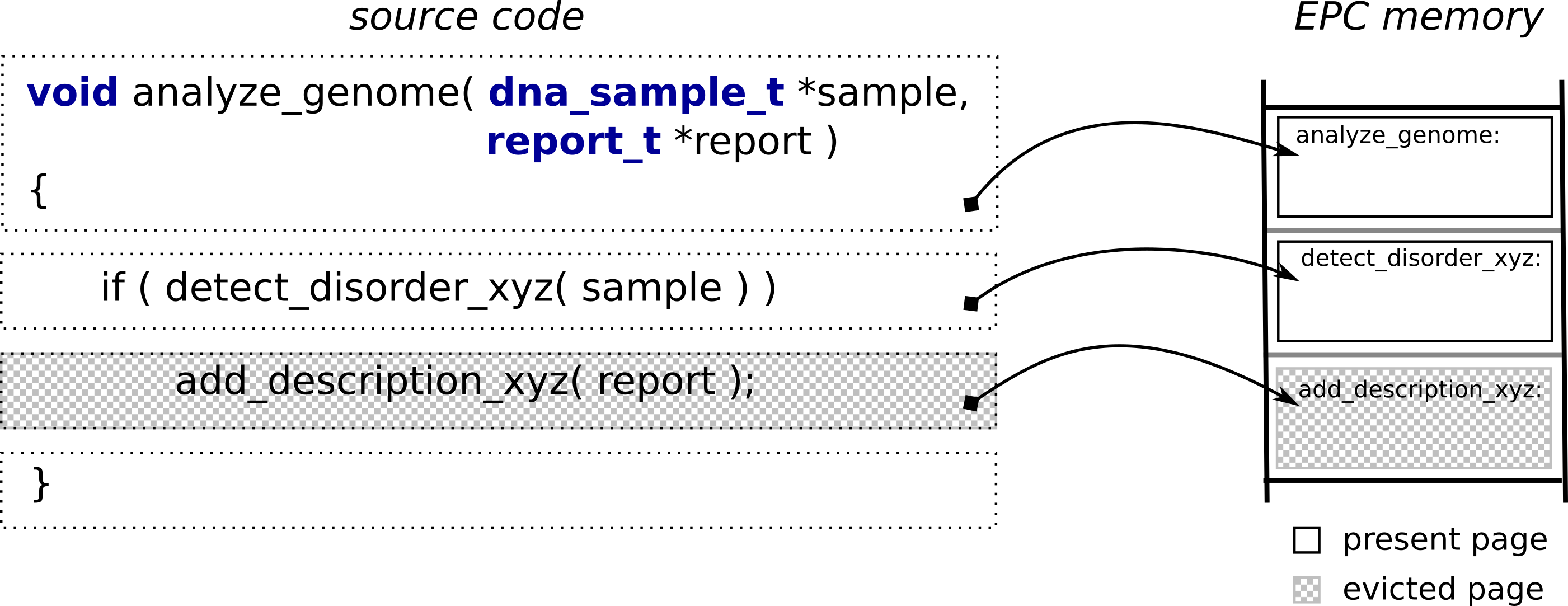}
    \caption{Basic example of a 
page-table based side channel attack. The occurrence (or absence) of a page fault 
after an attacker evicted the
{\tt add\_description\_xyz} function
reveals the result of the DNA analysis.}
    \label{fig:basicAttack}
\end{figure}

\paragraph{Basic Attack}
\label{sec:basicAttack}
Consider as an example the code shown in Figure~\ref{fig:basicAttack} that
detects and reports the presence of specific mutations in a patient's genome. 
To ensure the confidentiality of patients' medical records, the analysis is
placed in an enclave. 

In the most basic form of the attack, the page containing the 
{\tt add\_description\_xyz} function is evicted from EPC memory. 
When during enclave execution a page fault is raised,
the function was called and the attacker learns that the patient suffers from
the specific disorder. 
Note that also the absence of a page fault leaks information about the patient's health. 


This very basic example assumes that the {\tt add\_description\_xyz}
function is located on an otherwise unused page. 
Such requirements can be eliminated~\cite{xu2015controlledChannel} when key locations of the 
enclave's call graph are evicted. 

\paragraph{Closely-related Attack Vectors}
%


At Usenix'17~\cite{vanbulck2017pfless} and CCS'17~\cite{wang2017leakyCauldron} even more 
subtle attacks were demonstrated:

\begin{itemize}
 \item An enclave page in EPC memory but marked non-present in
the page tables will also lead to a page 
fault.~\cite{xu2015controlledChannel,weichbrodt2016asyncshock}
 \item Writing to an enclave page marked non-writable in the page table leads to a page fault,
 even when EPCM metadata indicates write privileges. Revoking 
 execution rights from code pages leads to similar events.

 \item Accessed and dirty bits are still recorded in the page 
 tables for EPC memory~\cite{vanbulck2017pfless,costan2016sgxExplained,wang2017leakyCauldron}. 
 \item Page table entries are also placed in the data cache when they are accessed.
 Hence, an attacker monitoring the cache can detect when the page table is walked 
 and which enclave page is accessed.~\cite{vanbulck2017pfless,gras2017aslr,wang2017leakyCauldron}
\end{itemize}

Especially these last two attacks are hard to defend against as they can be executed
concurrently from another logical core {\em without} causing an enclave exit.
This significantly reduces the effectiveness of state-of-the-art 
defenses~\cite{chen2017dejavu,shih2017tsgx}.

\subsection{Attack Model}
To provide strong protection against the various ways enclave-page accesses
are revealed, we assume a strong attack model where an attacker can
(1) observe {\em all} page table walks in enclave mode {\em without} 
interrupting the enclave's execution~\cite{vanbulck2017pfless,wang2017leakyCauldron} and (2) interrupt an
enclave precisely at instruction-level 
granularity~\cite{vanbulck2017pfless,hahnel2017highResolution}.

Although other side channels exist that threaten the confidentiality of 
enclave execution~\cite{lee2017branchShadowing}, we consider those to be out of scope. In 
particular, we do not consider cache-attacks~\cite{gotzfried2017sgxCache,
schwarz2017sgxCache,brasser2017sgxCache,moghimi2017cache}
against EPC memory. This is
an orthogonal problem. Any protection against an attacker probing an enclave's cache, 
still needs to consider the possibility that 
enclave pages may be evicted from EPC memory and its page tables manipulated or observed. 

\subsection{Security Guarantees}
\label{sec:securityGuarantees}
\label{sec:discussionReactiveApproaches}

Most state-of-the-art solutions that protect SGX enclaves against page-table-based
side channels are reactive security measures. They try to
detect when the enclave is under attack based on unusual numbers of 
discontinued code execution~\cite{shih2017tsgx} or time taken to execute security sensitive 
code~\cite{chen2017dejavu}. Such approaches are suboptimal as (1) novel attacks 
do not cause enclave exits and cannot be observed from the enclave itself~\cite{vanbulck2017pfless}, 
(2) the absence of 
page faults may also leak sensitive information to an attacker (see 
Section~\ref{sec:basicAttack}) and (3) enclaves when not under
attack may also experience page faults
(e.g., a benign OS paged out an enclave page) and interrupts (e.g.,
an external interrupt needs to be handled). Hence it is challenging 
for enclave developers to 
specify a threshold of enclave exits that balances the likelihood that 
attacks are being stopped with a low number of false-positives (see Appendix~\ref{app:evaluation:tsx}). 
(4) Reactive defenses require special security policies~\cite{shih2017tsgx,chen2017dejavu}
to ensure that after the detection of an attack, enclaves can be restarted securely. Attackers 
must not be able to launch attacks 
against different instances of the same enclave and aggregate the 
information learned from each individual attack. This is important especially in
server settings where services should always be up 
and running. Moreover, while cryptographic keys can often be re-negotiated~\cite{chen2017dejavu}, 
it remains unclear how security policies can prevent attacks against 
the sensitive data itself that an enclave processes (e.g., the genome data
in Section~\ref{sec:basicAttack}).

Heisenberg in contrast takes a completely different approach that avoids
these shortcomings. Instead of detecting
an ongoing attack, we proactively take security measures to ensure that
any attack against the enclave will either be detected {\em before} any confidential data 
is processed, or fail to observe enclave page accesses. 
An enclave protected with Heisenberg will thus never leak even a single
bit of information about its internal state to an attacker through its page accesses.
Hence, there is no need to differentiate benign from malicious interrupts; 
enclaves may {\em always} be resumed. 

\subsection{Operational Guarantees for System Software and Enclaves}
\label{sec:operationalGuarantees}





Any solution to the page-table-based side channel problem, needs to ensure that
all operational properties of system software and enclaves can still be guaranteed. 

First, it is paramount that system software remains in full control over {\em all} 
system resources, including EPC memory. The operating system and virtual machine
monitor (VMM) must be able to efficiently and fairly multiplex EPC memory. This
implies that (1) enclave pages must {\em always} be able to be evicted from EPC 
memory, (2) the hardware should aid the EPC multiplexer to pick good EPC page 
candidates to be evicted~\cite{costan2015sanctum}, for example through the 
accessed/dirty bit mechanism
in its page tables and (3) enclaves should not be able to
hold a lock on EPC memory~\cite{shinde2016pidgeonHoles} nor the 
CPU (i.e., interrupts should not be disabled~\cite{volp2016enclaveSidePreemptionControl}).
This is especially challenging in a cloud setting as EPC resource allocation 
by the VMM should also be transparent to the virtual machine's OS managing the same
resource. Complex negotiation mechanisms between enclave and system software 
are an intractable solution.


Second, when the system is {\em not} under attack, correctly-written 
enclaves should never end up in a state where they cannot advance.

\section{The Heisenberg Defense}
\label{sec:heisenberg}





Earlier designs of protected-module architectures have been implemented through a 
more privileged software layer 
such as a hypervisor~\cite{mccune2010TrustVisor,strackx2012fides} 
or system management mode (SMM) \cite{azab2011sice}. Whenever a context switch is performed between the 
untrusted context and the protected module, this privileged layer
enforces the memory access control mechanism. This is also a prime 
location to implement 
defenses against various types of side channels. 
Hypervisor-based architectures for example could virtualize EPC memory and 
completely hide page faults from the untrusted virtual machine, foiling Xu et~al.'s 
attack~\cite{xu2015controlledChannel}.

Intel SGX provides very similar features. 
Unfortunately as they are implemented
almost completely in microcode~\cite{costan2016sgxExplained}, their flexibility is limited. 
Virtualizing EPC memory through microcode, for example, is infeasible as this implies accessing
disk space.

%


We propose a generic approach to bridge this gap between HW/SW-based approaches. 
Instead of a more privileged layer implementing defenses against side-channels, we enable
SGX enclaves to implement defensive measures upon enclave entry/re-entry. This has the advantage
that enclaves can use detailed information about their own implementation to reduce overhead.

To implement this approach, two requirements need to be met:

\noindent {\bf Requirement 1: Hookable security code}: An SGX enclave should always be able 
to execute defensive 
code inside the enclave before confidential data is processed. Upon enclave entry this is
trivial as enclaves always enter at a fixed set of locations. Enclave reentries on the other hand
pose technical challenges as SGX does not natively support intercepting their execution.



\noindent {\bf Requirement 2: Unobservable enclaved execution}: After the defensive code 
finished execution,
the enclave's behavior must not be observable by an attacker. This cloaking
must be done pro-actively as 
operational guarantees prevents us from (temporarily) disabling or postponing 
interrupt handling.

In Section~\ref{sec:basicDefense} we present our basic defense strategy. 
Heisenberg-HW (Section~\ref{sec:heisenbergHW}) provides complete protection 
at the cost of only very limited hardware modifications. Heisenberg-SW (Section~\ref{sec:heisenbergSW})
on the other hand can be applied on readily available platforms,
but the use of Intel TSX makes performance overhead hard to estimate.




%

\subsection{Heisenberg's Basic Defense Mechanism}
\label{sec:basicDefense}

\begin{figure}
    \centering
    \includegraphics[width=0.45\textwidth]{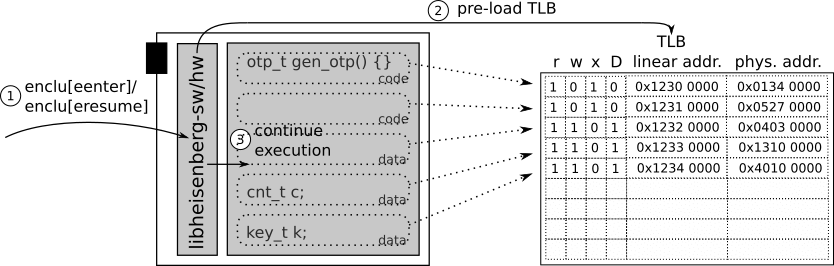}
    \caption{An attacker is prevented from observing memory access patterns by 
    intercepting all enclave (re-)entries and 
    pre-loading the TLB before sensitive data is processed.}
    \label{fig:heisenbergOverview}
\end{figure}

Consider an SGX enclave implementing a one-time password generator~\cite{hoekstra2013sgxUseCases}.
Figure~\ref{fig:heisenbergOverview} displays our approach graphically. When 
the enclave is entered or resumed, we intercept the control flow. For every
enclave page whose accesses may be of interest to an attacker, 
we load its address translation in the TLB. Only afterwards code to generate
the next password is executed. An attacker may monitor the page 
tables, but she will only observe pre-determined page accesses by our proactive loading scheme.
During password generation, address translations are already present in the TLB and no 
sensitive information leaks.
Interrupting the enclave would flush the TLB,
but the hooked {\tt libheisenberg-sw/hw} library would again pre-load the TLB before 
resuming the actual password generator.

To avoid all page table accesses and page faults during OTP generation, we 
issue dummy instructions. First, we verify that the enclave page is present in EPC memory,
marked present in the page tables and has the proper access rights.
For read-only enclave pages, it is sufficient to issue simple 
read instructions to these pages.
For writable pages we verify write access by reading and re-writing a single memory location. 
Executable enclave pages are first
searched for a return instruction and subsequently called.

Second, we ensure that during OTP generation, 
page table entries are no longer accessed to set accessed and dirty bits. 
As the page table is walked only when access to the page 
is requested, the accessed bit is set immediately. 
Accessed bits are not stored explicitly in the TLB; entries themselves indicate that
the bit is already set. The value of the dirty bit
on the other hand is cached in the TLB and set upon the first time a write
is issued. Both accessed and dirty bits are set only
once. An attacker clearing the bit from another core without flushing
the TLB, will not cause additional page table walks.~\cite{intel64manuals}

Finally we ensure that an attacker cannot evict a TLB entry during the 
execution of security-sensitive code. This is handled differently depending on 
whether we rely on hardware modifications.
We discuss TLB and EPC memory size-constraints in detail in Section~\ref{sec:discussion:enclaveSizes}.

\subsection{Heisenberg-HW: Pro-Active Code with Limited Hardware Modifications}
\label{sec:hwSupport}
\label{sec:heisenbergHW}




When considering hardware modifications to increase security, two research directions exists. 
First, one could reconsider the complete platform architecture and design an inherently more 
secure platform. This is interesting from a research perspective and valuable for new real-world products,
but it is unlikely that existing platforms will adopt these changes. Hardware vendors need to remain
backwards compatible with previous platform generations and new features will need to be supported 
almost indefinitely. 

We choose the other approach: bringing the desired features to the platform with only very limited, 
local changes. These changes are much less likely to affect existing software and are much more 
likely to be added to existing platforms.

\subsubsection{Requirement 1: Hooking {\tt libheisenberg-HW}}
Our first requirement is to guarantee that pro-active security code will always
be called within the enclave upon every enclave (re-)entry. This is non-trivial
as naive implementations may cause enclaves to enter a stuck state, even under 
benign conditions.


\paragraph{Operational Considerations}
At design-time enclave writers need to reserve a fixed stack of SSA frames.
When an enclave is interrupted, the top SSA frame is used to store the current
state of the processor. {\tt enclu[eresume]}s simply pop the top SSA frame
and resume execution.
The {\tt enclu[eenter]} instruction on the other hand checks that at least one slot
is still available on the stack, should the enclave call be interrupted.

A stack of SSA frames is needed when enclaves internally may cause a 
fault that cannot be handled by system software (e.g., div-by-zero).
In such situations another entry point can
be {\tt enclu[eenter]}'d to resolve the fault. Afterwards
the in-enclave fault handler exits the enclave and the subsequent
{\tt enclu[eresume]} resumes execution using the modified enclave context.

The question that now arises is how big the SSA stack should be. When 
enclave resumes would be handled in software, these functions themselves could be
interrupted before the SSA stack is popped, and this ad infinitum. This would quickly consume
all available SSA frames, prevent {\tt enclu[eenter]} instructions 
from succeeding and thus leaving the enclave in a stuck state.
By providing a hardware mechanism that handles resume-from-interrupt as an 
atomic {\tt enclu[eresume]}
instruction, SGX guarantees that under benign events correctly-written enclaves
will never end up in a state where they cannot advance.

\paragraph{Limited Hardware Modifications}
To provide hardware support to hook interrupts without limiting SGX' operational 
guarantees, we advocate an approach similar to the way faults are handled. 
We propose the implementation of a {\tt block-eresume} bit in SSA frames. 
Enclave writers who need to intercept interrupts, can set this bit in the SSA
frame on top of the SSA stack; the SSA frame that will be used to store
the enclave's state upon the next interrupt.


When an {\tt enclu[eresume]} is issued to resume an enclave, 
the {\tt block-eresume} bit of the SSA frame to be resumed from
is checked. When the bit is clear (the default), the enclave resumes execution as usual. 
When this bit is set however, the {\tt enclu[eresume]}'s fails. 
Similar to faults, the surrounding application must first call an 
interrupt handler within the enclave to reset the {\tt enclu[eresume]} bit.
As the enclave context (i.e., the SSA frame) used to be resume execution from 
can be modified as well at this point,
{\tt enclu[eresume]} instructions can be hooked .



\paragraph{Hooking Pro-Active Code with HW Support}
With these hardware changes, it is straightforward to force the execution of a
{\tt resume\_hook} function before an enclave resumes its normal operation. 
\Listing{lst:heisenbergHWhook} displays the algorithm. When the {\tt ecall\_gen\_otp} enclave 
entry point is called, we signal that future enclave interrupts need to be 
blocked (line~\ref{lst:heisenbergHWhook@genOtpBlockResume}). 
Next, {\tt proactive\_security\_code} is called which will (eventually) pre-load 
the required TLB entries. Finally the {\tt gen\_otp} function
is executed protected from page-table-based side channels 
and we clear the {\tt block\_eresume} bit and exit the enclave. 

When an interrupt arrives when the {\tt block-eresume} bit has been set, it can 
only be resumed again after its {\tt interrupt\_handler} was called. Here the enclave's
context (i.e., it's SSA frame) is temporarily stored in a 
global variable (line~\ref{lst:heisenbergHWhook@storeContext}). 
Next, the SSA frame is overwritten with an artificial {\tt hook\_context} 
to redirect execution to the {\tt resume\_hook} function and finally the SSA's {\tt block\_eresume}
bit is cleared. 
For clarity we postpone the discuss the meaning of the condition test on 
line~\ref{lst:heisenbergHWhook@flatten}.

When a {\tt enclu[eresume]} instruction is now issued, the enclave resumes execution
from the the {\tt hook\_context}. Again we ensure that all interrupts will force the execution of 
the {\tt interrupt\_handler} and execute {\tt proactive\_security\_code}. Finally we 
resume the enclave state. 

A subtle difficulty arises when we consider interrupts arriving while the {\tt resume\_hook} 
is being executed. Following the same principle we could hook such events but this would 
(1) require us to store a stack of hook contexts to be resumed from. As the {\tt resume\_hook}
function could be interrupted indefinitely, an infinite amount of hook contexts would
need to be reserved. (2) The {\tt proactive\_security\_code} is only valuable when 
executed (completely) before the enclave context is resumed. Resuming this function 
after it has been interrupted, only leads to performance overheads. 
Both problems are resolved by ensuring that when 
{\tt resume\_hook}'s execution is interrupted, its state is {\em not} stored but
the complete function is re-started from line~\ref{lst:heisenbergHWhook@resumeHookInst1}. 
We achieve this using a boolean
variable {\tt in\_hook} to keep track whether we're currently executing the {\tt resume\_hook}
function. Unfortunately resetting this variable after the enclave's context has been restored 
is infeasible. Alternatively, clearing the variable before restoring the context causes the 
{\tt enclave\_context} to be overwritten when an interrupt arrives between both instructions.
We resolve this problem by adding another condition that tests whether the instruction pointer
points within the {\tt resume\_hook} function. Only when both conditions indicate that we're 
outside the {\tt resume\_hook} function, will the {\tt enclave\_context} be overwritten 
(line~\ref{lst:heisenbergHWhook@flatten}).

\begin{lstlisting}[language=C,floatplacement=tb,float,caption={By adding a {\tt block-eresume}
bit in SSA frames, {\tt enclu[eresume]} instructions can be easily intercepted.},
label={lst:heisenbergHWhook},
otherkeywords={ob_group_t,ssa_t,entry_point,hb_frame_t,otp_t,sgx_status_t,byte,bool}]
ssa_t enclave_context;
byte hook_stack[0x1000];
bool in_hook;

sgx_status_t entry_point ecall_gen_otp( otp_t *otp ) {
  ssa_t *ssa = ssa_stack.top();
  ssa->block_eresume = 1; //* \label{lst:heisenbergHWhook@genOtpBlockResume} *//
  if ( proactive_security_code() != SGX_SUCCESS ) // e.g., preload_tlb() //* \label{lst:heisenbergHWhook@securityCode} *//
    return SGX_FAILED;
  *otp = gen_otp();
  ssa->block_eresume = 0;
  return SGX_SUCCESS;
}

sgx_status_t entry_point interrupt_handler( void ) {
  ssa_t *ssa = ssa_stack.top().prev();
  ssa_t hook_context;
  
  if ( ssa->block_eresume != 1 ) //* \label{lst:heisenbergHWhook@interruptFetchSSA} *//
    return SGX_FAILED;
  
  if ( in_hook == 0 && in_resume_hook( ssa->rip ) == 0 ) //* \label{lst:heisenbergHWhook@flatten} *//
    memcpy( &enclave_context, ssa, sizeof( ssa_t ) );  //* \label{lst:heisenbergHWhook@storeContext} *//
  
  in_hook = 1;
  hook_context.rip = &resume_hook;
  hook_context.rsp = &hook_stack[0x1000];
  memcpy( ssa, &hook_context, sizeof( ssa_t ) );
  ssa->block_eresume = 0;
  return SGX_SUCCESS;
}

void resume_hook( void ) { //* \label{lst:heisenbergHWhook@resumeHook} *//
  ssa_t *ssa = ssa_stack.top(); //* \label{lst:heisenbergHWhook@resumeHookInst1} *//

  ssa->block_eresume = 1;  //* \label{lst:heisenbergHWhook@setBlockEresumeInHook} *//

  if ( proactive_security_code() != SGX_SUCCESS ) // e.g., preload_tlb() //* \label{lst:heisenbergHWhook@securityCode2} *//
     simulate_aex();
  
  in_hook = 0;
  restore_from_context( &enclave_context ); //* \label{lst:heisenbergHWhook@resumeGenOtp} *//
}
\end{lstlisting}

\subsubsection{Requirement 2: Unobservable Enclaved Execution}
The introduction of a {\tt block-eresume} bit enables us to preload the TLB after 
every enclave (re-)entry. Unfortunately, this is not sufficient to defend against page-table-based
attacks. When HyperThreading is available and enabled, the TLB is shared between both logical cores.
Recent work~\cite{wang2017leakyCauldron} showed that an attacker may leverage this design choice 
and evict TLB entries of the enclave while executing on the other logical core. 

Various options to avoid such situations exists: the TLB could be split 50/50 between logical
cores, the TLB eviction policy could be changed, etc. Unfortunately, these changes would also
affect legacy code relying on this implementation. Therefore, we choose another approach
and verify that both logical cores have executed a {\tt rendez\_vous} function and 
are still executing within the enclave.

To implement the approach, we rely only on the following hardware modifications within SGX. First,
an SGX enclave is able to register for interrupts of the entire {\em physical} core. When HyperThreading
is enabled, interrupts, faults and traps are only received by one of the logical cores. We update 
this behavior so that when the other logical core is interrupted, this immediately results in an 
asynchronous enclave exit as well. Afterwards the interrupt is handled as usual by the logical 
core it was send to.

Second, when the enclave on the other logical core executes an {\tt enclu[eexit]} instruction,
the enclave should experience an AEX.

Finally, the enclave is able to determine the id of the physical core it is currently running on. 
The {\tt rdtscp} instruction already returns the id of the logical core it is executed on, but
the instruction is illegal in SGXv1. In SGXv2 it is unclear whether its results can be 
spoofed by a hypervisor.

\Listing{lst:heisenbergHWrendezvous} displays our protocol. As it needs to rendezvous with the 
other logical core every time the enclave is (re-)entered, we re-use the hooking feature of the 
previous section. In fact,
\Listing{lst:heisenbergHWrendezvous}, line~\ref{lst:heisenbergHWrendezvous@proactiveCode} displays
the code called at \listing{lst:heisenbergHWhook} lines~\ref{lst:heisenbergHWhook@securityCode} 
and \ref{lst:heisenbergHWhook@securityCode2}.

As the first step, the {\tt rendez\_vous} protocol registers itself 
(line~\ref{lst:heisenbergHWrendezvous@registerInterrupt}) to receive an AEX whenever
either logical core receives an interrupt, trap or fault. 
When we also verify that both logical cores execute within the same enclave, both
will continue their enclaved execution until one of them exits. When that happens, 
both will (eventually) re-execute the {\tt rendez\_vous} function.

For convenience and according to standard practices in 
mutual exclusion algorithms, we assume that the variable {\tt i} always 
refers to the logical core that the code is running on (i.e., 0 or 1). The variable {\tt j} 
refers to the other logical core ({\tt j = (i + 1) mod 2}). 

Our rendez-vous protocol works by having both logical cores write the id of their physical
core to two shared variables {\tt id[i]} and {\tt id[j]}. When both match they 
execute within the same enclave and on the same physical core. A {\tt mutex} lock prevents
concurrent access by both logical cores.
But care needs to be taken. The hooking mechanism of \Listing{lst:heisenbergHWhook} only ensures
that the code will be re-executed after an interrupt. The value of shared variables (e.g., {\tt id})
are left unchanged and may be stale. Similarly, once a lock is taken, it may not be
released again.

We resolve this locking problem by recording who currently holds the lock; the logical core requesting
the lock (i.e., {\tt i}), the other logical (i.e., {\tt j}), or whether the lock is free (i.e., {\tt -1}).
To take the lock we use a compare and exchange instruction {\tt bool cmpxchg(\&lock, v, w)}. 
This hardware instruction is readily available in x86 processors. It guarantees that {\tt true} is returned
and the value of the lock changed to {\tt w} {\em iff} {\tt lock} has the current value {\tt v}. Otherwise
the value of the lock remains unchanged. The instruction is guaranteed to operate atomically.
\Listing{lst:heisenbergHWrendezvous}~line~\ref{lst:heisenbergHWrendezvous@getLock} ensures that
a lock is only taken when it is free or the logical core already received it.


Once the lock is taken, we erase the physical id recorded by the other logical core 
(i.e., {\tt id[j]}) and set our own identifier (i.e., {\tt id[i]}) in 
lines~\ref{lst:heisenbergHWrendezvous@eraseID} and \ref{lst:heisenbergHWrendezvous@setID}.
As long as the other logical core does not catch up and records its physical id -- possibly 
for the second time -- we enter a wait loop (lines~\ref{lst:heisenbergHWrendezvous@forLoop} 
and \ref{lst:heisenbergHWrendezvous@rerun}).
Finally we check whether both logical cores execute on the same physical core 
(line~\ref{lst:heisenbergHWrendezvous@testID}).

The protocol works by always first erasing the physical id of the other logical core in the 
first iteration of the {\tt for} loop. All other iterations, its own physical id is (re-)written
to {\tt id[i]}. This implies that when the conditions on line~\ref{lst:heisenbergHWrendezvous@rerun}
evaluates to {\tt true}, {\em the logical core {\tt j} {\em must} have executed after the {\tt id[j]} was
erased}. As this is set only within the {\tt rendez\_vous} function at 
line~\ref{lst:heisenbergHWrendezvous@setID}, both logical cores are executing within the enclave and 
at the same time.

\begin{lstlisting}[language=C,floatplacement=tb!,float,caption={Rendez-vous protocol ensuring that both logical cores have entered the same enclave.},
label={lst:heisenbergHWrendezvous},
otherkeywords={ob_group_t,ssa_t,entry_point,hb_frame_t,otp_t,shared,sgx_status_t,atomic}]
shared int mutex = -1;
shared int id[2] = {-1, -1};

sgx_status_t proactive_security_code( void ) { //* \label{lst:heisenbergHWrendezvous@proactiveCode} *//
  if ( rendez_vous() != SGX_SUCCESS )
    return SGX_FAILED;
  preload_tlb();
  return SGX_SUCCESS;
}

sgx_status_t rendez_vous( void ) {
  // register for interrupt/abort of this physical core
  register_interrupt( read_physical_core_id() ); //* \label{lst:heisenbergHWrendezvous@registerInterrupt} *//
  
  for ( int iteration = 0; true; ) { // busy waiting for rendez-vous //* \label{lst:heisenbergHWrendezvous@forLoop} *//
    // take free or already owned lock
    while( !cmpxchg( &mutex, -1, i ) && !cmpxchg( &mutex, i, i ) ){} //* \label{lst:heisenbergHWrendezvous@getLock} *//
    
    switch ( iteration ) {
      case 0:
        id[j] = -1; // always erase the other one's variables //* \label{lst:heisenbergHWrendezvous@eraseID} *//
        ++iteration;
      default:
        id[i] = read_physical_core_id(); //* \label{lst:heisenbergHWrendezvous@setID} *//
        
        if ( id[j] != -1 ) { // wait for other core to catch up //* \label{lst:heisenbergHWrendezvous@rerun} *//
          if ( id[i] != id[j] ) { //* \label{lst:heisenbergHWrendezvous@testID} *//
            mutex = -1;
            return NOT_ON_SAME_PHYSICAL_CORE;
          }
          else
            return RENDEZVOUS_SUCCESS;
        }
    }
    
    mutex = -1; // release lock
  } // end for-loop
}
\end{lstlisting}

\subsection{Heisenberg-SW: A Software-Only Approach}
\label{sec:tsxApproach}
\label{sec:heisenbergSW}

While Heisenberg-HW only relies on very limited hardware modifications,
they are not yet readily available. We build a software-only version of
our approach by relying on Intel TSX. 



\subsubsection{Requirement 1: Hooking {\tt libheisenberg-SW}}
%

The use of Intel TSX is straightforward.
When we issue an {\tt xbegin~$a$} instruction, a new transaction is started.
From then on all 
memory accesses remain within the processor's cache, hidden
from other actors on the platform. 
Upon a memory conflict, interrupt, etc.~the transaction is aborted,
intermediate results are discarded from the cache and
execution is resumed from memory location {\tt $a$}.
Only when an {\tt xend} instruction
is executed, is the current transaction committed. 

Consider \listing{lst:heisenbergInterface} 
line~\ref{lst:heisenbergInterface@rtmbegin}. In this code snippet
the {\tt xbegin} instruction is wrapped in a function. When
the {\tt rtm\_begin} function is called, a new transaction is started and 
the value {\tt 0} is returned. However, when the 
transaction is aborted in the future, the entire system state is 
rolled back and execution resumes from line \ref{lst:heisenbergInterface@abortPoint}.
The application appears to have returned {\tt -1} immediately when {\tt rtm\_begin}
was called.

\paragraph{Hooking {\tt enclu[eresume]}s}


Transactions abort when an interrupt or fault occurs. We use this feature
to intercept enclave execution flow after the {\tt enclu[eresume]} 
was issued. We 
do so by wrapping all confidentiality sensitive code executed by an enclave in a transaction. 
\Listing{lst:heisenbergInterface} displays an example of an enclave generating 
a one-time password.
When the enclave is called to generate a new password 
(\listing{lst:heisenbergInterface}, line~\ref{lst:heisenbergInterface@entryPoint}), 
a new transaction  is started.
Next, all security sensitive enclave pages are preloaded in the TLB, as 
specified in Section~\ref{sec:basicDefense}.
Afterwards normal enclave execution is resumed and an OTP password is generated.

When an interrupt arrives, the transaction is aborted and the 
enclave is exited immediately. Only after the interrupt is serviced,
is the enclave re-entered and the RTM transaction abort handler called 
(\listing{lst:heisenbergInterface}, line~\ref{lst:heisenbergInterface@abortPoint}). Eventually
this leads to preloading all security sensitive pages again and the re-execution 
of the entire 
transaction (\listing{lst:heisenbergInterface}, line~\ref{lst:heisenbergInterface@whileLoop}).
Before the enclave returns to unprotected memory,
the transaction is committed.


To account that transactions are constrained both in space and time,
we commit intermediate results to main memory (line~\ref{lst:heisenbergInterface@commit}). 
However, executing an {\tt xend} and {\tt xbegin} sequentially is not sufficient. 
An attacker interrupting the newly started transaction, 
will automatically flush the TLB. When the enclave eventually resumes its 
execution, an attacker could again learn the enclave's 
page accesses by monitoring its page table walks. 
We prevent such attacks by forcing the pages to be preloaded again
whenever a new transaction is started.

\begin{lstlisting}[language=C,floatplacement=tb!,float,
caption={Heisengberg-SW confidentiality sensitive enclave code in RTM transactions
and reuses the transaction abort handler to intercept enclave resumes after interrupts.},label={lst:heisenbergInterface},
otherkeywords={ob_group_t,entry_point,otp_t,sgx_status_t,asm}]
sgx_status_t entry_point ecall_gen_otp( otp_t *otp ) { //* \label{lst:heisenbergInterface@entryPoint} *// 
  while ( rtm_begin() == -1 )//* \label{lst:heisenbergInterface@whileLoop} *//
    ; // retry the transaction 
  
  preload_tlb();
  *otp = gen_otp();
  asm volatile ( "xend\n" );
  return SGX_SUCCESS;
}

int rtm_begin( void ) {//* \label{lst:heisenbergInterface@rtmbegin} *//
  int ret;
  asm volatile ( "  xbegin 1f"  	//* \label{lst:heisenbergInterface@xbegin} *//
               "  mov $0, %0"
               "1:"  
               "  mov $-1, %0"    //* \label{lst:heisenbergInterface@abortPoint} *//
               : "r"(ret) :: );
  return ret;
}

void heisenberg_commit( void ) { //* \label{lst:heisenbergInterface@commit} *//
  asm volatile ( "xend\n" );
  while ( rtm_begin() == -1 )
    ; // retry the transaction 
  
  preload_tlb();
}
\end{lstlisting}

\subsubsection{Requirement 2: Unobservable Enclaved Execution}
Out of the 189 SGX-capable processors released by Intel,
66 do {\em not} support HyperThreading~\cite{intel2017search}. 
For these processors this requirement
is fulfilled automatically. 
For the others care needs to be taken to prevent an
attacker from evicting TLB entries from another core~\cite{wang2017leakyCauldron}.

For processors that do support HyperThreading, preloading TLB entries still significantly
raises the bar for an attacker. As the TLBs replacement policy has not yet
been reversed engineered, leaking enough information from a running enclave is non-trivial.

Stronger security guarantees can be provided by relying on secure boot.
At boot time PCR registers are extended with a measurement of all code and configuration
parameters the system booted from. A secure channel can be set up between the enclave and 
the TPM chips that holds the PCR registers. This enables enclaves to verify that HyperThreading
has been disabled. Setting up an authenticated session with the TPM chip and reading PCR values 
is estimated by 
related work at 72ms~\cite{strackx2016ariadne}. The performance can be 
improved significantly by the introduction of a proxy enclave. As HyperThreading cannot be
re-enabled after the BIOS disabled it, the proxy enclave could query the TPM chip only once per 
power cycle. All other enclaves can set up a secure connection with the proxy enclave to request
the result of its last PCR reading. 

\section{Implementation}
\label{sec:implementation}

We implemented a prototype for both proposed defenses. As we cannot modify x86 microcode,
the Heisenberg-HW prototype assumes a co-operating runtime that calls the 
{\tt interrupt\_handler} enclave entrypoint whenever the enclave exits asynchronously. While 
this cannot provide any security guarantees against a malicious application or OS, it does
enable us to run performance benchmarks on existing hardware. We also modified the 
Intel sgx\_sign tool to store the enclave's layout data (e.g., page access rights) in 
enclave-protected read-only memory. sgx\_edger8r was modified to hook calls to the enclave. 
Outward calls  to unprotected memory are hooked similarly. Table~\ref{tbl:codeSizes} displays
the extend of our code modifications.



The Heisenberg-SW does {\em not} rely on a co-operating runtime system.
At enclave compile-time, we wrap sensitive code in an TSX transaction.
There exists a delicate balance 
on the size of transactions. Large transactions that are aborted (e.g., due 
to an interrupt) will lead to the loss of a large amount of intermediate results
that will need to be re-computed. Smaller transactions reduce the amount of 
results lost, but require that every enclave page is re-visited
whenever a new transaction is started. 
We implemented a simple but effective heuristic that roughly keeps track of how many
instructions have been executed within the transaction. When a certain limit 
is reached, the transaction is committed.

Our LLVM pass implementing the heuristic takes two arguments: {\tt FUNC\_SKP}
and {\tt INIT\_CNTR}. Instrumenting small functions would make
for a more accurate estimate, but the induced performance overhead for 
such functions can be significant. Therefore we ignore functions smaller
than {\tt FUNC\_SKP} instructions (in LLVM's IR language).
In larger functions runtime code is added to the function's prologue. 
A global {\tt counter} is used to track the number of LLVM IR instructions
that can still be executed within the current transaction. When the function is
called, we decrement {\tt counter} with the size of the current function. When
{\tt counter} drops below zero, the current transaction is committed, and 
a new one started. The {\tt counter} variable is reset to {\tt INIT\_CNTR}.
A similar approach could be applied to loops as well. Although this would provide
better estimates of the transaction size, it would also increase the number of 
instructions on the enclave's control flow. We decided not to implement this
heuristic.

%

\begin{table}
\begin{center}
\begin{tabular}{lrrrr}
               & Lang. & Vanilla & Added & Total \\ \hline 
\multicolumn{5}{|c|}{Heisenberg-HW} \\ \hline
\multirow{2}{*}{libheisenberg-hw} & \multirow{2}{*}{$\left\{\begin{tabular}{@{\ }l@{}}
     C\\ asm 
  \end{tabular}\right.$}  & - & 102 & 102 \\
               &  - &  -& 207 & 207 \\ \hline
\multicolumn{5}{|c|}{Heisenberg-SW} \\ \hline
\multirow{2}{*}{libheisenberg-sw} & \multirow{2}{*}{$\left\{\begin{tabular}{@{\ }l@{}}
     C\\ asm 
  \end{tabular}\right.$}  & - & 243 & 243 \\
               &  - &  -& 207 & 207 \\
LLVM Pass     &  C++ & - & 148 & 148 \\ \hline
\multicolumn{5}{|c|}{Shared library/tools modifications} \\ \hline
sign\_tool    &  C++ & 2,742        &  102         &  2,844      \\
\multirow{2}{*}{urts}  & \multirow{2}{*}{$\left\{\begin{tabular}{@{\ }l@{}}
     cpp\\ asm 
  \end{tabular}\right.$}  & 3,979 & 33 & 4,012 \\
               &  - & 146 &  23 & 169 \\

edger8r & ML & 1,615 & 38 & 1,653\\
\end{tabular}
\caption{Code Sizes for {\tt libheisenberg-sw/hw} and its compilation and runtime support.}
\label{tbl:codeSizes}
\end{center}
\end{table}

\section{Evaluation}
\label{sec:evaluation}
\subsection{Operational Evaluation}

It is paramount for any defense for Intel SGX not to limit system software
in its abilities to govern access to platform resources. Neither Heisenberg-SW 
nor Heisenberg-HW limit the untrusted kernel or hypervisor 
from regaining execution control. At the moment an interrupt is raised, 
enclaves immediately exit asynchronously. 

In addition neither modify the behavior nor meaning of the 
accessed and dirty bits in the page tables. Hence, system software can still
rely on these bits to manage EPC memory, including when any optimization 
is implemented as proposed in Section~\ref{sec:discussion:enclaveSizes}. Enclaves
may require enclave pages to be present in EPC memory before any enclaved computation 
is performed, but cannot cause deterioration of the operation of other enclaves. 
System software remains in charge and 
can opt to reduce CPU time for enclaves requiring too much EPC memory. 
Commodity systems already take similar defensive measures against applications
consuming too much main memory causing the system to start trashing.~\cite{denning1968trashing}

Finally Heisenberg-HW ensures that correctly written enclaves do 
not get in a stuck-state under benign events. Once the enclave is prevented
from being resumed, it is up to the enclave writer to ensure that this block
can be undone. As by default SSA frames are always resumable, a fixed
size for the SSA stack can always be found. Heisenberg-SW does not modify 
the hardware and thus trivially fulfills this requirement.

\subsection{Security Evaluation}
\label{sec:evaluation:security}

Our Heisenberg defense relies on the fact that we can intercept all 
enclave calls and resumes, and essentially ``blind'' an attacker 
before executing security sensitive code.
We distinguish between platform and compilation considerations.

\paragraph{Platform Considerations}
An attacker targeting an enclave protected with Heisenberg can attempt
several types of attacks. Any attack that relies on the interruption
of the enclave will fail. After preloading all enclave pages that may be used
before security-sensitive data is used in the enclave, no information 
leaves the TLB. When an interrupt arrives, the enclave is exited, and 
the TLB flushed automatically as specified by the SGX 
documentation.~\cite{intel64manuals}

Attacks that rely on incorrect access rights or the inspection of the page tables and/or its caches also
fail: Pre-loading enclave pages in the TLB guarantees that access rights have not been 
reduced and their accessed and 
dirty bits are set. Until their entries are evicted from the TLB, no accesses
to the page tables in memory will be made, avoiding information leaks.
Information may leak to an attacker during the TLB preloading phase, 
but as all pages are accessed in a pre-determined order, no sensitive information leaks.

Finally all attacks during enclave execution are prevented. As no 
information leaks after the TLB is preloaded, an attacker can only attempt 
to flush entries. Several options exist. 
Instructions such as move-to-{\tt cr3} and {\tt invlpg} flush TLB entries directly, 
but can only be executed on the 
same logical core as the affected TLB. Issuing such instructions would thus first entail
an interrupt to unprotected (kernel) memory and no sensitive information can be learned. 

Enclave pages could be evicted from EPC from another core. 
As SGX already relies on the correctness of TLB entries during enclave execution,
such events could also affect the integrity of enclave execution. To prevent 
such events, SGX hardware already ensures that pages are evicted only after
all affected logical cores received an inter-processor-interrupt (IPI), 
flushing their TLB.~\cite[\S 38.5.3]{intel64manuals}


We also need to ensure that all preloaded page translations fit in the TLB. For the 
Skylake microarchitecture~\cite[\S 2.1.3]{intel64optimization} 
1,536 L2 TLB entries are available
on the physical core with an associativity of 12.
Unfortunately, to the best of 
our knowledge, the algorithm behind this TLB entry assignment has not yet been reversed
engineered. However, given the locality principle, adjacent pages will most likely
be mapped to different TLB cache sets. We discuss the impact of enclave and TLB sizes in detail
in Section~\ref{sec:discussion:enclaveSizes}.

\paragraph{Enclave Considerations}

Enclaved code also needs to avoid that TLB entries are 
evicted inadvertently. Defensive measures need to be taken during enclave development.
First, we need to ensure that after enclave pages are preloaded, no
accesses to unprotected memory are made. An attacker that can force 
another address translation to be inserted in the TLB, may cause 
an enclave-page entry to be evicted. 
Such events can be avoided by preloading TLB entries
 {\em after} data is copied from unprotected memory to the enclave.

Second, in case of Heisenberg-SW we also need to consider special 
instructions (e.g., {\tt enclu[egetkey]}) that cause TSX transactions to 
abort. At compilation-time we can detect
such calls and avoid that they reveal any sensitive information. For 
example by buffering the enclave's seal key during
the first call to the enclave. 

\subsection{Performance Evaluation}



We ran our experiments on a standard Dell Desktop equipped with an 
Intel Skylake i7-6700 processor, 16 GiB RAM (no swap space) and a 
SATA-connected SSD. During the benchmarks the machine was connected
on an ethernet network. We also disabled SpeedStep, P-States,
C-states and TurboBoost to get more stable benchmark results.

\subsubsection{Microbenchmarks}
\label{sec:evaluation:microbenchmarks}

\begin{figure}[tb]
    \centering
    \includegraphics[width=0.4\textwidth]{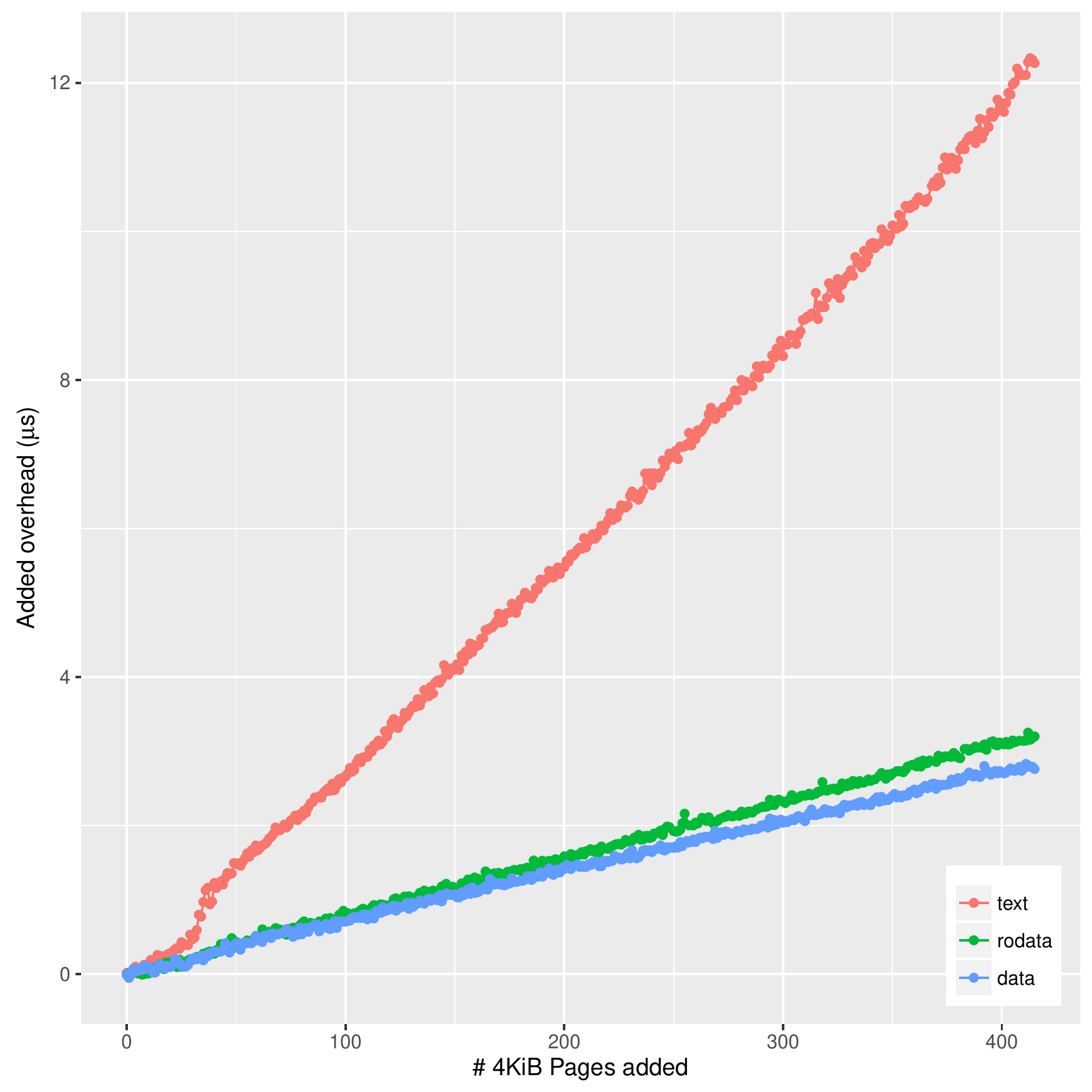}
    \caption{Overhead of increasing enclave sizes per page type.}
    \label{fig:tlbFill}
\end{figure}

\paragraph{Impact of Filling the TLB}
To assess the impact of filling the TLB, we created a dummy enclave with
an empty enclave function. We called the enclave 1,000 times and recorded
the required time. Next, we gradually increased the number of read-only,
writable and executable pages. The results are displayed in Figure~\ref{fig:tlbFill}.

As expected adding an executable page is the most costly at 0.030~$\mu$s per page.
More surprising was that adding a read-only page is with 0.008~$\mu$s/page 
more time-consuming than adding a writable page (0.007~$\mu$s/page). 

Unfortunately, when running the benchmark within a TSX transaction we see that when we 
added 316 additional data pages (totaling 325 pages), we 
reached a hard limit. To ensure that we have write access to data pages, we need
to read/write to every page. This quickly consumes a large fraction of the transaction's 
write set. After adding 316 pages, insufficient space is left to execute the microbenchmark.
We discuss the impact of enclave sizes in Section~\ref{sec:discussion:enclaveSizes}.

\subsubsection{Macrobenchmarks}
\paragraph{Executing on an Idle System}


Benchmarking applications that wrap large code sizes in TSX transactions 
is challenging. Implementing an application and measuring its median execution
time over a large number of runs, does not reliably indicate how the application
will perform in a real-life scenario. In Section~\ref{app:evaluation:tsx} we will
show that performance is impacted by (1) whether the previous transaction succeeded. 
Once a transaction succeeds, it is much more likely to run to completion again. This
favors standard benchmark settings, but in practice an application may only be
executed at large time intervals. (2) What the load is on the system in a multicore 
processor, even when we reserve a core for our application and offload IRQs. 

Even though we acknowledge these shortcomings, we implemented two benchmarks
and execute them on a reserved (physical) core. 
Table~\ref{tbl:smallBenchmarks} displays the results. 
First, we implemented a small enclave that recursively calculates 
the $40^{th}$ Fibonacci number and measured its performance over 100 runs. 
Executing the enclave protected with Heisenberg-SW ({\tt INIT\_CNTR}=1,800, {\tt FUNC\_SKP}=1)
requires over 2~million intermediate commits leading to a performance overhead of 22.54\%.
When protected with Heisenberg-HW, the enclave was interrupted 173 times over a 
median time of 681.42ms. This resulted in a speedup of 4.57\%. We suspect that by 
preloading TLB-entries, the out-of-order execution engine performs much better as
pipeline stalls due to page-table walks are reduced.

While Shinde~et~al.~\cite{shinde2016pidgeonHoles} indicate that their 
SHA-512 implementation confines input-dependent memory accesses to a single page,
this may not always be the case. Linking additional code in the same enclave, may 
spill memory accesses to adjacent pages. Nor is this sufficient. Van~Bulck~et~al.~\cite{vanbulck2017pfless}
show that even such enclaves can be attacked successfully. As expected, protecting
our SHA-512 implemenation ported from Arm Mbed~\cite{armmbed} with Heisenberg-HW finishes
in only 10.24$\mu$s. In all 1,000 runs the execution was not interrupted and 
overhead remains negligible with only 1.56\%. When protecting the enclave with
Heisenberg-SW, performance improves significantly (34.42\%). Again we believe this is caused
by the processor's cache replacement strategy that attempts not to abort TSX transactions
due to cache conflicts in the (processor-wide) L3 cache.
As transactions need to remain in all cache levels (i.e., the L3 cache is 
inclusive) until completion and each SHA-512 invokation will reference the same cache lines, 
it will suffer from very low cache pressure. This is consistent with 
the results presented in Appendix~\ref{app:evaluation:tsx:reexecution} \S1. 
The Heisenberg-HW and unprotected enclave
versions, do not experience this favorable cache replacement strategy.
%
%

\begin{table}
\small
\fontsize{7pt}{8pt}\selectfont
\begin{tabular}{l|r|rr|rrrrr}
\hline
   & unprot. & \multicolumn{2}{|c|}{Heisenberg-HW} & \multicolumn{3}{|c}{Heisenberg-SW} \\ \hline
Benchmark                     & time & overhead & \# int. &   overhead & \# aborts & \# commits \\ \hline
Fibo                          & 714.052ms & -4.57\% & 173 & 22.54\% & 1,544, & 2,019,273 \\
SHA512                        & 10.087$\mu{}$s & 1.56\% & 0 & -34.43\% &    0   & 5\\
\hline
\end{tabular} 
\caption{Microbenchmarks for Heisenberg-HW/-SW when executing on an idle system.}
\label{tbl:smallBenchmarks}
\end{table}

\paragraph{Executing under Realistic System Load}
We also ported the FreeType library version 2.5.3 to bare SGX hardware. 
The same version was attacked by Xu~\cite{xu2015controlledChannel} and proved 
too complex for Shinde et~al.~\cite{shinde2016pidgeonHoles} to defend against 
using a compiler-based defense. FreeType provides
a type setting service. In our benchmark, we type set ``Hello World!'' under 
an 25\textdegree{} angle on a canvas of 640 by 480 pixels. Each pixel was 
recorded as a single character. 

\begin{table*}
\small
\fontsize{7pt}{8pt}\selectfont
\begin{tabular}{l|r|rrr|rrrrrr}
\hline
   & unprotected & \multicolumn{3}{|c|}{Heisenberg-HW} & \multicolumn{6}{|c}{Heisenberg-SW} \\ \hline
System load                     & time (in ms) & time (in ms) & overhead & \# re-entries &   time (in ms) & overhead & {\tt INIT\_CNTR} & {\tt FUNC\_SKP} & med aborts & \# commits \\ \hline
idle                            & 1.247 & 1.256 & 0.66\% & 0                               &    7.357 & 489.85\% & 1,100 & 6 &  5 & 2,240 \\
idle                            & 1.247 & 1.256 & 0.66\% & 0                               &    {\bf 4.978} & {\bf 299.11\%} & {\bf 1,100} & {\bf 11} &  {\bf 13} & {\bf 2,189} \\ \hline
writes                          & 1.249 & 1.261 & 0.96\% & 0                               &    7.967 & 538.05\% & 1,100 & 6 & 166 & 2,240 \\ 
writes                          & 1.249 & 1.261 & 0.96\% & 0                               &    {\bf 5.826} & {\bf 366.52\%} & {\bf 1,100} & {\bf 11} & {\bf 192} & {\bf 2,189} \\ \hline
I/O stress                      & 1.259 & 1.320 & 4.88\% & 0                               &    9.180 & 629.16\% & 1,100 & 6 & 846 & 2,240 \\ 
I/O stress                      & 1.259 & 1.320 & 4.88\% & 0                               &    {\bf 6.238} & {\bf 395.46\%} & {\bf 1,100} & {\bf 11} & {\bf 538} & {\bf 2,189} \\ \hline
llvm                            & 1.288 & 1.369 & 8.38  & 0                               &    {\bf 7.366} & {\bf 471.94\%} & {\bf 1,100} & {\bf 6}  & {\bf 9} & {\bf 2,240} \\ 
llvm                            & 1.288 & 1.369 & 8.38  & 0                               &    41.826 & 3,147.39\% & 1,100 & 11 & 10,777 & 2,189 \\ \hline
\hline
\end{tabular} 
\caption{Benchmark results for FreeType under various conditions and 
settings. Figures in bold shows Heisenberg-SW's configuration with the smallest overhead under the given system load.}
\label{tbl:freetypeBenchmark}
\end{table*}

With 109,046 lines of C code, this enclave has a considerable size. 
As we also included the Arial font in the enclave, the overall 
enclave size is also much larger with 439 pages in total. Unfortunately,
this proved to be too large for Heisenberg-SW. We could not write-access all 107 data pages, and 
resorted to only read-accessing them.
The access rights of executable pages were verified correctly. 

We tested 80 different Heisenberg-SW configurations on an idle system with {\tt INIT\_CNTR} values 
ranging between 100 and 2,000 (steps of 100) and {\tt FUNC\_SKP} ranging between 1 and 16 (steps of 5). 
The best performing configuration of Heisenberg-SW resulted in an overhead 
of 299.11\% ({\tt INIT\_CNTR}=1,100, {\tt FUNC\_SKP}=11) with 2,189 intermediate 
commits (see Table~\ref{tbl:freetypeBenchmark}). 

As Intel TSX performance depends heavily on system load (see Appendix~\ref{app:evaluation:tsx}),
we performed the same benchmark while the other available cores were indefinitely writing 
to a 2~MiB buffer (``writes''), performing disk accesses with the {\tt stress}-tool and 
compiling the llvm-3.9 compiler. Even though we reserved a physical core for our benchmarks 
and interrupts were offloaded to another core, the number of aborts skyrocketed from 13 (idle) to
10,777 (llvm). We could not reduce overhead by searching for a better configuration for Heisenberg-SW, 
with one notable execution. During llvm compilation performance overhead could be reduced 
from 3,147\% to 471\% by also inlining functions larger than 6 LLVM IR LoCs instead of 11. 
But as the median number of aborts also dropped from 10,777 to only 9, we expect this is caused 
by variations in system load.

Heisenberg-HW performs much better when tested under the same system loads; overhead 
increased only from 0.66\% to 8.38\% as more than 50\% of enclave calls were not interrupted
and did not require pre-loading the TLB again. 
This strengthens our believe that IRQs {\em can} be offloaded to different
cores, but that this has only limited impact for TSX transactions.

\section{Discussion: Enclave Sizes}
\label{sec:discussion:enclaveSizes}

The root cause of Xu's et~al.~\cite{xu2015controlledChannel} and related 
attacks~\cite{vanbulck2017pfless,wang2017leakyCauldron} is that EPC memory is limited.
In current implementations only 128~MiB RAM memory is reserved for SGX of which
only 96~MiB can be allocated to store
enclaved code and data. Many SGX use cases remain well below this limit,
requiring only a few hundred kilobytes to a few megabytes of EPC memory. This
includes software implementations of hardware-based security solutions~\cite{hoekstra2013sgxUseCases}, 
security sensitive parts of 
applications~\cite{mccune2008Flicker,mccune2010TrustVisor,strackx2012fides,
singaravelu2006ReducingTCB}, and 
enclaves that provide a security service to untrusted 
applications~\cite{mccune2009SafePassage,strackx2014ice,strackx2016ariadne}.

When multiple small enclaves are present in the system 96~MiB of EPC memory may not be sufficient.
This is especially challenging in a cloud setting where a hypervisor may virtualize EPC memory,
oblivious to virtual machines running on top; a kernel cannot guarantee
that all enclave pages are present when an enclave is entered or resumed. 
For previously proposed solutions~\cite{chen2017dejavu,shih2017tsgx} such conditions will 
trigger a false-positive conclusion that the enclave is under attack.
Heisenberg's reactive approach handles such situations more gracefully. An enclave page that has been
paged out by the hypervisor will result in the page fault when 
the TLB is pre-filled. After the enclave page is loaded back in EPC memory, 
execution can be resumed safely.

In Skylake processors only 1,536 TLB entries~\cite{intel64optimization} are available. 
Enclaves larger than 6~MiB
exceed this limit and their page address translations cannot completely be preloaded. 
In such cases we must resort to static analysis. Attackers observing the page table
can only learn enclave secrets when secret dependent page accesses occur. But many
pages are accessed independent of sensitive information. Such page translations
can be omitted during Heisenberg's TLB pre-loading phase. Gruss~et~al.~\cite{gruss2017tsx} take such
an approach to defend against cache attacks.
Note that TLB pressure would be reduced 512x when support for 2~MiB pages is added 
to SGX, avoiding any need for static analysis. 

Enclaves larger than 96~MiB are challenging. As they are always partially
paged out of EPC memory, they always reveal some information. 
Probabilistic defenses such as ASLR~\cite{seo2017sgxShield} handle such situations
gracefully, but only increase the bar for page-table-based attacks 
slightly~\cite{xu2015controlledChannel,vanbulck2017pfless,wang2017leakyCauldron}.
We propose two future research directions. General-purpose solutions for large
enclaves could apply mechanisms from oblivious RAM (ORAM) implementations to
hide page accesses, not individual memory accesses. 
Liu~et~al.~\cite{liu2015ghostRider} propose an interesting technique 
by applying ``scratchpad'' memory that cannot be observed by an attacker. Heisenberg 
could provide scratchpad memory whose page accesses are oblivious to an attacker. 

Alternatively, special-purpose solutions can be applied. 
Fuhry~et~al.~\cite{fuhry2017hardIDX} consider the problem of enclaves that 
search in a huge data set and need to hide 
access patterns to their search tree. Their solution places the 
integrity and confidentiality protected search tree
in unprotected memory.
An attacker observing accesses to this tree will not
learn any information. 
Such solutions still require that page accesses of the enclave itself
are oblivious to an attacker.

\section{Related Work}
\label{sec:relatedWork}
Academic researchers have been proposing protected-module architectures for a long 
time~\cite{mccune2008Flicker,singaravelu2006ReducingTCB}.
Surprisingly page-table-based side channels against such architectures have only 
been described~\cite{xu2015controlledChannel,vanbulck2017pfless,wang2017leakyCauldron} 
recently. However, some early designs are 
not vulnerable to such attacks 
as they set up their own page tables~\cite{mccune2008Flicker},
avoid that page tables can be monitored~\cite{strackx2012fides,wilson2007trustzone} or simply 
because the targeted platforms do not provide virtual 
memory~\cite{noorman2013sancus,koeberl2014trustlite,brasser2015tytan}.

In Intel SGX page-table-based side channels cannot easily be avoided. As it also tries 
to defend against potentially malicious cloud 
providers~\cite{baumann2014haven,schuster2014vc3,brenner2016secureKeeper}, it cannot 
rely on a more privileged software layer such as a hypervisor. At the same time, it still wants to
enable system software to govern platform resources such as EPC memory. 
Designs where enclave pages are locked in main memory~\cite{ferraiuolo2015full} 
or where enclaves rely on referencing physical memory directly without 
a hypervisor controlling guest-physical to host-physical address page 
tables~\cite{evtyushkin2014isoX,costan2015sanctum}, are 
not acceptable.


Shih~et~al.~\cite{shih2017tsgx} propose in their NDSS'17 paper T-SGX, 
a reactive defense mechanism that wraps enclave execution in TSX transactions.
When the same transaction is aborted more than 5 times, this is seen as an attack
and the enclave is destroyed. 

Chen~et~al.\cite{chen2017dejavu} use a similar approach but rely on
Intel TSX in a separate enclave thread to construct a secure clock. 
The enclaved program checks the secure clock during its execution. 
Any significant deviation from the expected execution time, must have 
originated from an asynchronous enclave exit (AEX). Once detected
a security policy can be implemented to determine whether the system
is under attack and how it should be addressed.

Both T-SGX as D\'ej\`a-Vu suffer from limitations inherent to reactive 
defenses against page-table based side channels: (1) attacks may use 
information leaks that can be observed without causing enclave 
exits~\cite{vanbulck2017pfless}, (2) the absence of a page fault itself
already leaks information that a specific page is not required, 
(3) even under benign conditions enclaves will be interrupted, 
(4) once an attack is detected it is challenging to securely resume 
execution. In contrast, Heisenberg is a proactive defense mechanism 
that avoids these shortcomings (see Section~\ref{sec:securityGuarantees}).


At the SysTEX'16 workshop V\"{o}lp et~al.~\cite{volp2016enclaveSidePreemptionControl} proposed  
to enable SGX enclaves to temporarily disable interrupts in order to execute preparation code 
within the enclave. But this also requires a delicate balance.
A too short time frame would significantly limit the code that can be executed atomically.
A too large time frame limits
system software to handle interrupts in a timely manner and affects
fair scheduling of processes and virtual machines.

Other approaches rely heavily on rewriting software running in enclaves. 
Shinde~et~al.~\cite{shinde2016pidgeonHoles} implement a compiler pass that forces code
to be page-fault oblivious: by adding dummy branches it is ensured that all execution
traces access the same pages in the same order. It is assumed that once an enclave 
accesses a page, system software can only revoke access after a page fault. 
Interrupts~\cite{vanbulck2017pfless} are not considered. The authors also 
acknowledge that their defense technique cannot be applied when the enclaved application
(e.g., the FreeType library) follows a too unbalanced execution tree. 

Obfuscated enclave code make it harder to reconstruct or analyze control flow 
and also raises the bar for page-table based side-channel attacks.
Seo~et~al.~\cite{seo2017sgxShield} propose SGX-Shield, a fine-grained ASLR 
mechanism~\cite{bhatkar2003aslr} 
that randomizes 32~byte-sized execution units within the enclave in order 
to defend against low-level attacks against enclave code (e.g., ROP attacks~\cite{shacham2007rop}),
Schuster~et~al.\cite{schuster2014vc3} apply another approach
by loading and executing encrypted payloads. 
Both approaches only provide limited protection as (1) the decryption and 
randomization process itself could be attacked~\cite{xu2015controlledChannel}, 
(2) observing changes in 
page accesses given known enclave input/output may reveal valuable information.
For example an attacker targeting the example of Section~\ref{sec:basicAttack},
may learn that a specific page is accessed if and only if a specific disorder is detected,
by analyzing known DNA samples. Finally (3) 
publicly available libraries may exert unique page access footprints, 
potentially revealing which code is executing within the enclave. 
We investigate in future work.


%

Finally Intel suggests~\cite{intel2017SGXDevGuide} to
``align specific code and data blocks to exist entirely within a single page.''
The cryptographic community goes even further and proposes to eliminate branches 
completely~\cite{coppens2009practical}. 
While such solutions may be feasible for specific software, it is extremely hard 
for generic code and would lead to unacceptable performance overheads. 

Other related work tries to defend against other side channel attacks. 
Gruss et~al.~\cite{gruss2017tsx} for example showed that TSX could be used to cloak 
cache accesses by pre-loading data in transactions. But similar to Heisenberg-SW, this
approach may lead to significant performance overhead or even non-termination when 
applied on systems under heavy load. These disadvantages could be eliminated when 
our proposed hardware modifications are adopted for systems with an exclusive L3 cache (e.g.,
the Skylake-server architecture).

\section{Conclusion}
\label{sec:conclusion}
We presented Heisenberg, a proactive defense mechanism that provides complete
protection to enclaves against page-table based 
side channels. Heisenberg verifies and preloads enclave page translations
before confidentiality-sensitive code is executed. This enables it
to (1) protect against even the most novel page-table based attacks and 
(2) avoids the shortcomings of reactive defenses.

We evaluated two implementations. Heisenberg-SW uses
Intel TSX and can be readily applied on platforms lacking 
HyperThreading support (35\% of all SGX-capable processors). Platforms
with HyperThreading support can rely on PCR registers to determine whether
HyperThreading has been properly disabled in the BIOS.

We showed that Intel TSX can lead to huge performance overhead or even non-termination
when applied to hook enclave resumes. This makes any defense mechanism 
using Intel TSX in such a way a short-term solution. Heisenberg-HW 
relies on very limited hardware modifications to avoid these shortcomings.



\section*{Acknowledgment}
This research is partially funded by the Research Fund KU Leuven. 
Raoul Strackx is supported by
a grant of the Research Foundation - Flanders (FWO).
 
%



%
%
%

\bibliographystyle{plain}
\bibliography{papers_full}

\appendix
\section{Background: Intel Software Guard eXtensions}
\label{sec:background}



Some notable design decisions were made for Intel SGX to co-operate seamlessly 
with legacy operating systems. We introduce important aspects here,
but refer the reader to the Intel Manuals~\cite{intel64manuals} and 
other documentation~\cite{mckeen2013sgxInstructions,
anati2013sgxAttestationAndSealing,hoekstra2013sgxUseCases} for more details.

\paragraph{Basic Operation}
From a high overview, Intel SGX relies on two key properties: a 
program-counter-based access control mechanism and key derivation.
After modules are created, access rights change according to the 
currently executing instruction. Software running outside the enclave 
-- at any privilege level -- cannot access the enclave's memory region 
in any way. Only predefined entry points can be called. 
As these entry points are located within the enclave, memory regions
within the enclave become accessible. To enable enclaves to integrate
easily in OS processes, and to pass data to and from the enclave, (untrusted) 
process memory remains accessible.

This enclave isolation mechanism is only enforced after the enclave has been 
properly initialized. This opens up a small window of opportunity for an attacker
to modify the enclave when it is created. SGX' key derivation mechanism mitigates
such attacks. At creation time, a unique set of keys is derived based on a platform
secret and the initial state of the enclave. Any modification during set-up,
will lead to different cryptographic keys. This will prevent an attacker from (1) 
accessing sensitive data that was stored confidentiality and integrity protected by
a previous, genuine enclave and (2) attesting the execution of the modified enclave
to a third party.

\paragraph{Untrusted Page Tables}




SGX enclaves execute in the same address space as their process. 
This enables easy development of enclaves as data passed to and from enclaves does
not need to be marshaled. By simply passing a pointer to unprotected memory, 
the enclave itself can directly read and write in- and output. 


Even though the correct mapping of virtual enclave pages to physical memory is
paramount for security, the address translation units remain untrusted. Instead SGX applies
a use-but-verify approach. The untrusted legacy kernel remains in charge of
creating the segmentation and page tables, but the SGX hardware checks whether
this mapping is correct. 

When an logical address is referenced by a process, it is first passed through the
segmentation unit. 
As in x86 64-bit mode the base addresses and limits for segments {\%CS}, {\%DS}, {\%SS} and {\%ES} 
are enforced to be set to {\tt 0x00000000} and {\tt $2^{64}-1$} respectively,
no additional checks are performed. 

After the segmentation unit returned a linear address, it is passed through the 
paging unit. There the linear address is first located in the translation look-aside 
buffer (TLB). If it is not found there, the page tables (and extended page tables, when present) 
are walked until a physical
address is found eventually. Additional checks now need to be performed.
An attacker in control of the untrusted page tables, must not be able to map enclave pages
to incorrect physical memory or disable protection bits (e.g., W-xor-X prevention).
SGX facilitates this verification by dividing SGX reserved memory into an 
{\em enclave page cache} (EPC) and an
{\em enclave page cache map} (EPCM). Each 4~KiB page in EPC memory has a unique entry in the EPCM
that keeps track of the validity of the page, its type, a reference to which enclave it 
belongs to and so on. For regular pages it also keeps track of their linear
addresses they are supposed to be mapped on and their access rights. 
After the page table walk resulted in a physical location in EPC memory, its related EPCM entry
is fetched. When the metadata stored there matches with the translation of the page 
table unit, the linear-to-physical mapping is added to the TLB.
To ensure that the translations in the TLB remain up to date, 
the TLB is flushed automatically upon every enclave entry and exit.

\paragraph{Enclave Page Eviction}
Intel SGX does not only protect sensitive enclaved information from attackers who infiltrated
at any privilege level (kernel, hypervisor, etc.). 
It also protects against coldboot attacks~\cite{halderman2008coldBootAttacks} and
attackers snooping the memory bus. 
 A hardware encryption engine
ensures that enclave memory is only available in plaintext when it resides in one
of the CPU's caches. Before it is sent to main memory, it is on the fly confidentiality, 
integrity and version protected by the encryption engine. By statically allocating 
EPC/EPCM memory as a contiguous physical memory region of typically 128~MB, the 
implementation of the encryption engine becomes manageable. 

Unfortunately, this construct places a hard limit on the amount of protected memory. 
The overall demand for SGX memory may be much larger, especially in a cloud setting
where multiple clients wish to execute enclaves at the same time in their own virtual 
machines. 
The EPC/EPCM structures also play an 
important role here. Special instructions are available to system software to 
swap EPC pages in/out. A similar use-but-verify approach as for address translation
is used: system software selects which EPC page needs to be evicted from EPC memory and 
manages the swap space. SGX hardware ensures that these pages are properly protected
before leaving EPC memory and that the necessary checks are performed when they are
swapped back in.

\paragraph{Interrupts}
To ensure that system software remains in control over the platform, it is also
paramount that malicious or badly-behaving enclaves can be interrupted at any 
moment in time. To prevent that sensitive information may leak through such
interrupts, a stack of State Save Area (SSA) frames need to be reserved at enclave creation time. When
an interrupt is received, registers are stored in the top frame of the SSA stack. Registers
are then cleared and a dummy interrupt frame is pushed on the unprotected interrupt
stack of the kernel. When the 
operating system has serviced the interrupt, a legacy ``return-from-interrupt'' instruction 
({\tt iret}) is issued. This will result in the implicit calling to a 
resume-from-enclave-interrupt-handler
in the untrusted application. It is this untrusted handler that will issue the SGX
{\tt enclu[eresume]} instruction. Unlike legacy {\tt iret} instructions, 
an {\tt enclu[eresume]} will 
restore the entire processor state from the top SSA frame. It is not possible to
register an interrupt handler within the enclave.

%
%
%

\section{TSX Behavior under System Load}
\label{app:evaluation:tsx}

Intel TSX has been proposed to avoid the communication overhead originating from taking 
explicit locks in multithreaded applications. Unfortunately Intel TSX cannot
guarantee that transactions will ever be committed. To avoid such situations 
developers are urged~\cite{intel64optimization} to fall back on explicit synchronization 
locks when transactions
frequently abort. Unfortunately, this is infeasible when applied to hook
SGX resumes~\cite{shih2017tsgx}. We analyze Intel TSX' limitations.

\begin{figure}
    \centering
    \begin{subfigure}{0.23\textwidth}
        \includegraphics[width=\textwidth]{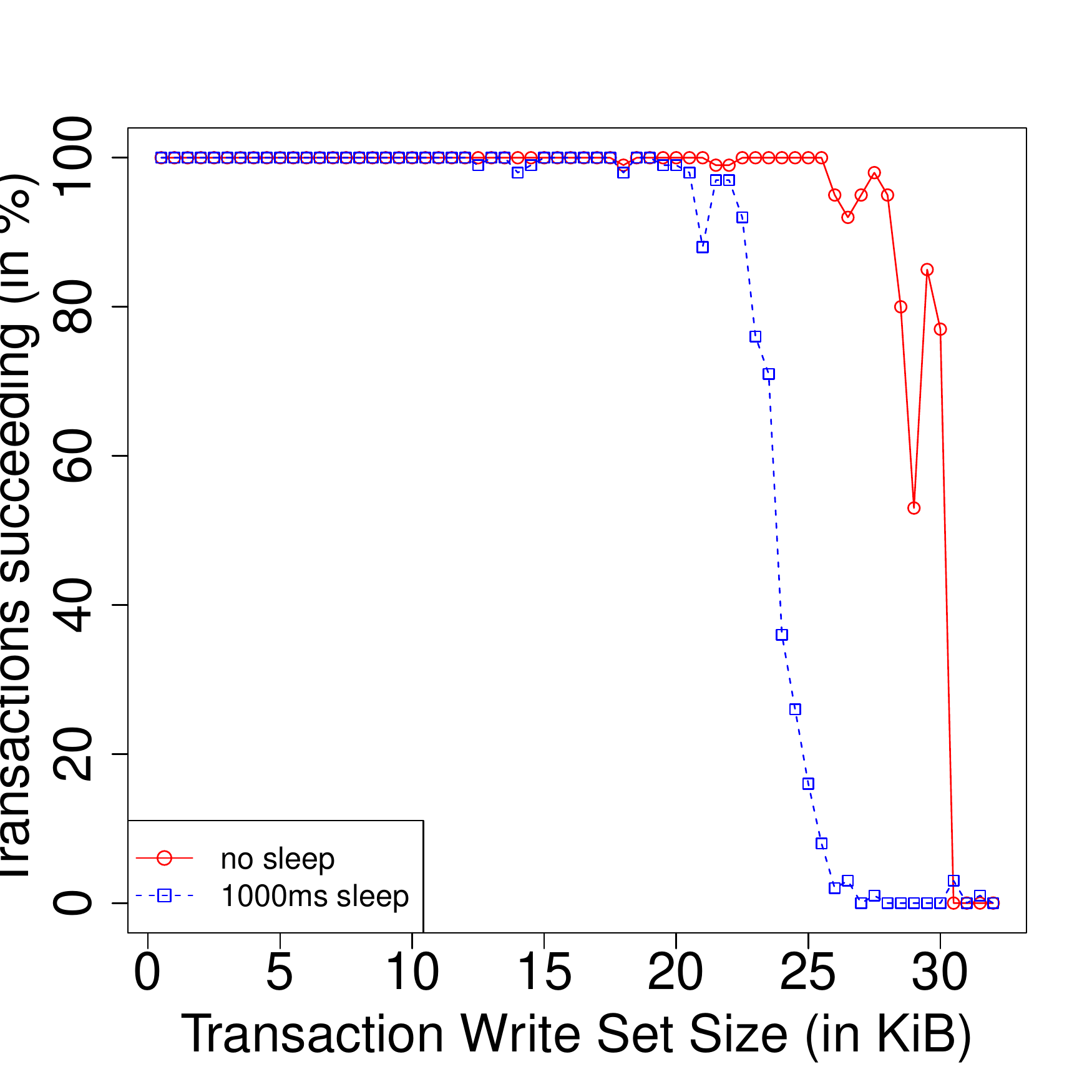}
        \caption{Write set size}
        \label{fig:rtmSuccesses:write}
    \end{subfigure}
    \begin{subfigure}{0.23\textwidth}
        \includegraphics[width=\textwidth]{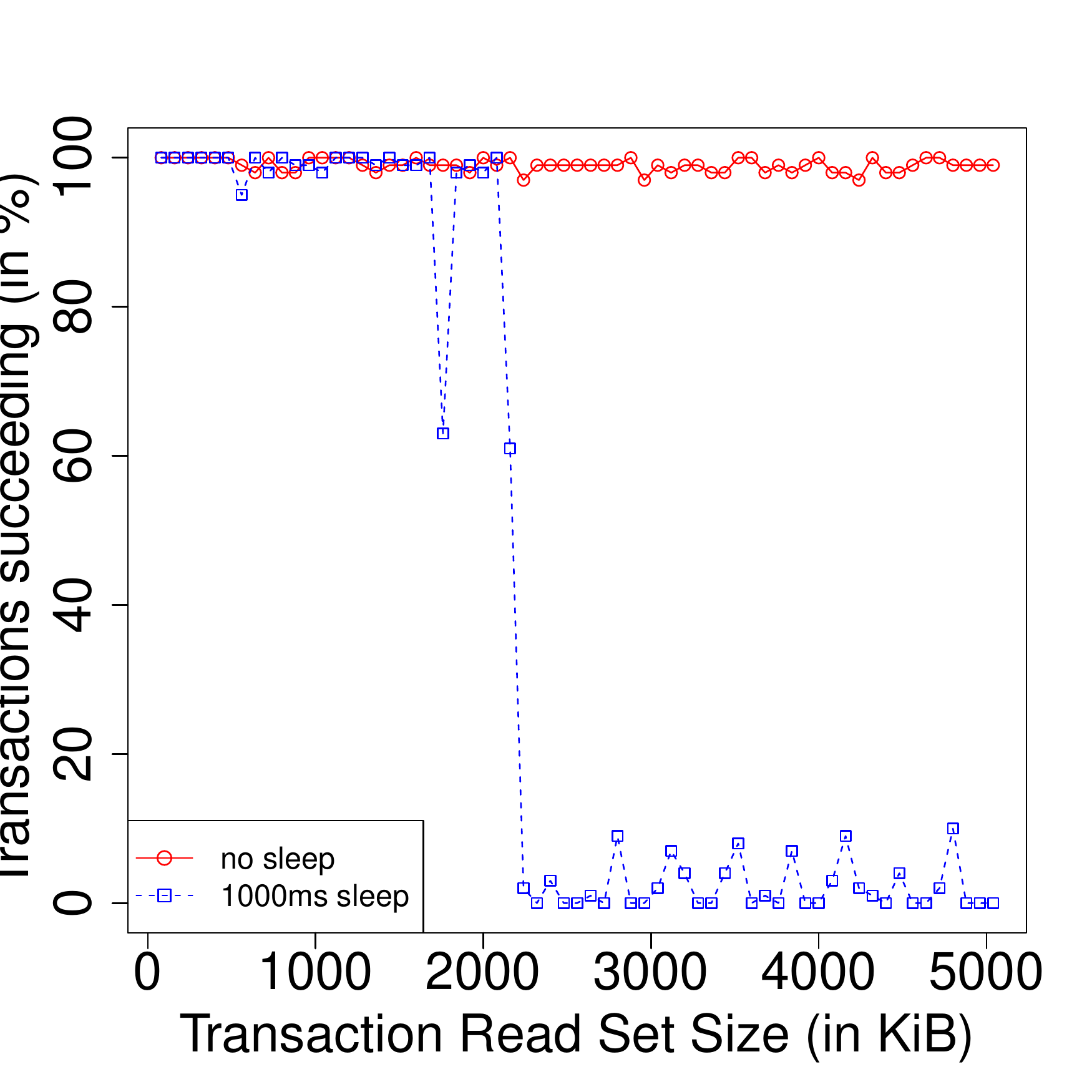}
        \caption{Read set size}
        \label{fig:rtmSuccesses:read}
    \end{subfigure}
    \caption{Test results of the size of RTM's read and write set.}
    \label{fig:rtmSuccesses}
\end{figure}


\begin{figure*}
    \centering
    \begin{subfigure}{0.45\textwidth}
        \includegraphics[width=\textwidth]{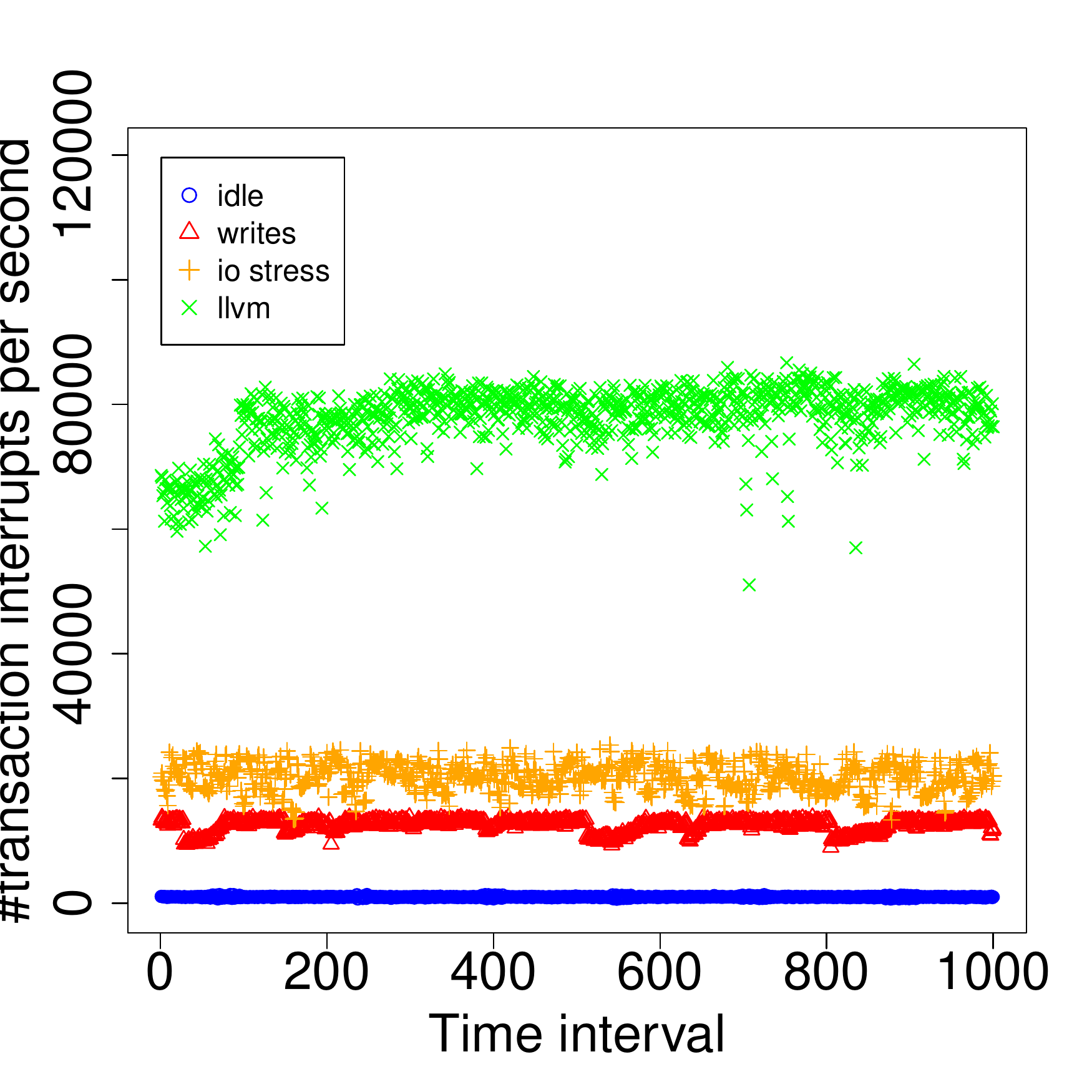}
        \caption{Executing {\tt while(1)\{\}} in a transaction on a reserved core still experiences 
        a large number of interrupts when the system becomes under heavy load.}
        \label{fig:rtmAborts:interrupts:whileTrue}
    \end{subfigure}
    ~\quad 
    \begin{subfigure}{0.45\textwidth}
        \includegraphics[width=\textwidth]{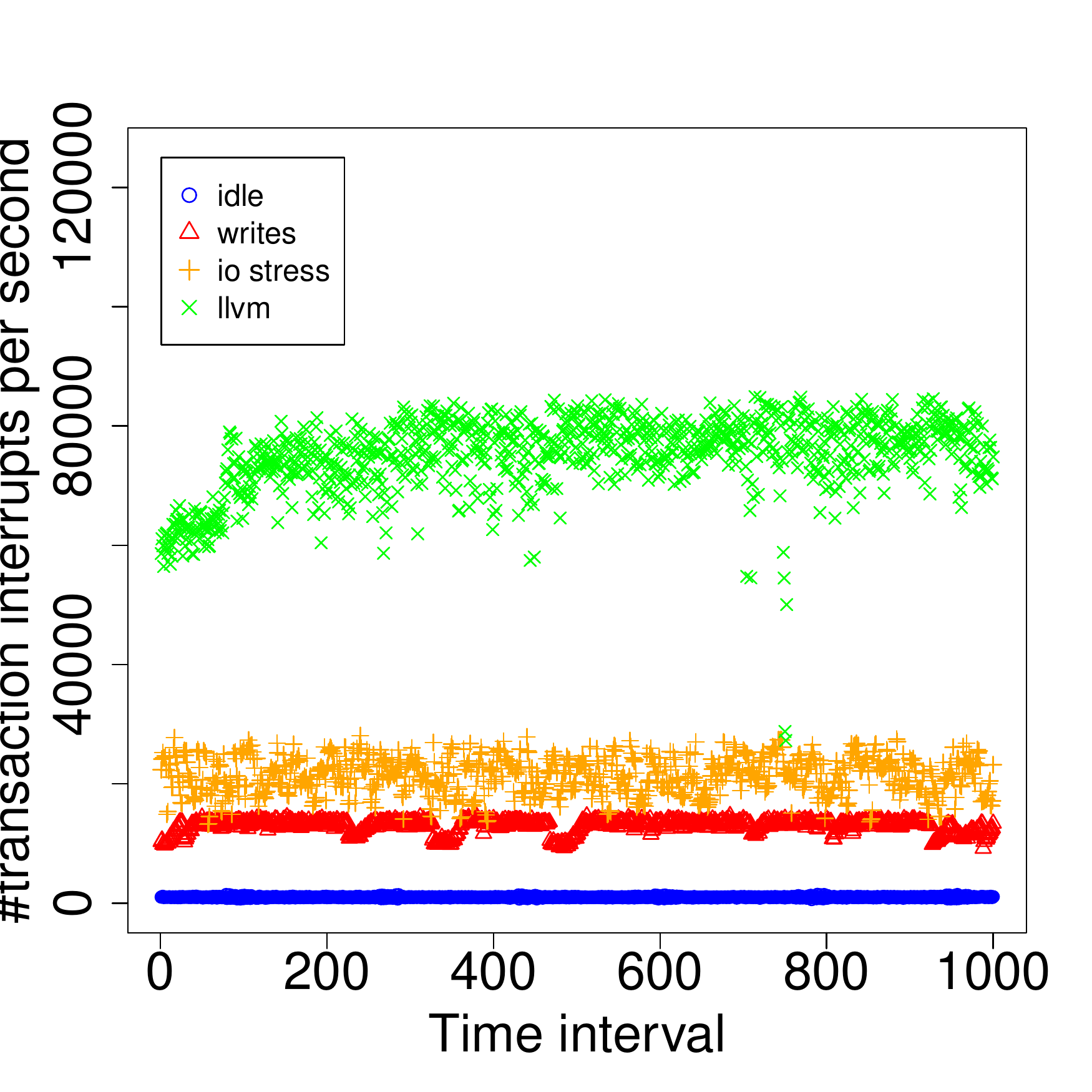}
        \caption{Reading 4 KiB in a transaction on a reserved core does not impact 
        the number of interrupts when the system becomes under heavy load.}
        \label{fig:rtmAborts:interrupts:read4K}
    \end{subfigure} \\
    
    \begin{subfigure}{0.45\textwidth}
        \includegraphics[width=\textwidth]{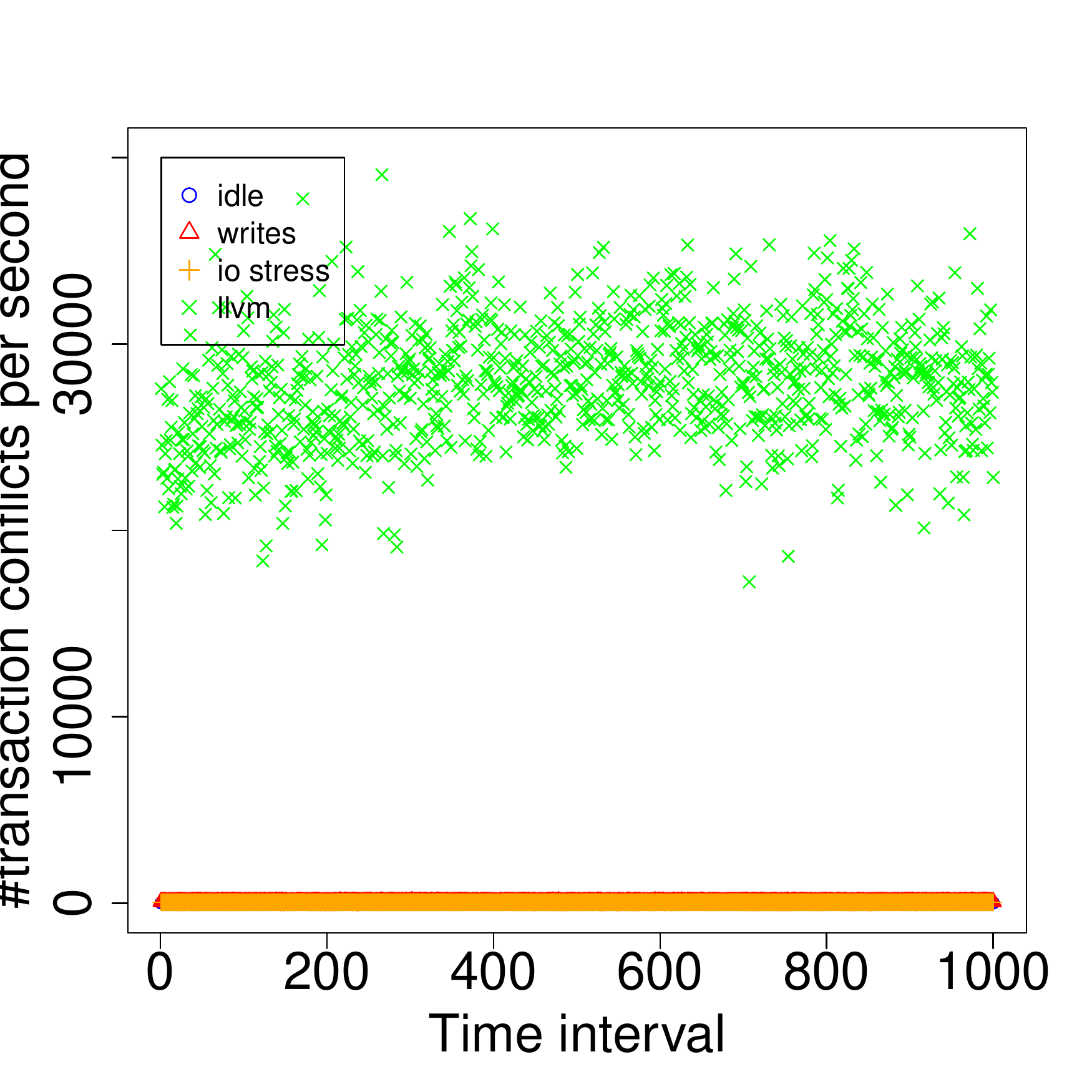}
        \caption{Executing {\tt while(1)\{\}} in a transaction on a reserved core results in many more 
        cache conflicts when the system comes under heavy load than an idle system.}
        \label{fig:rtmAborts:conflicts:whileTrue}
    \end{subfigure}
    ~\quad
    \begin{subfigure}{0.45\textwidth}
        \includegraphics[width=\textwidth]{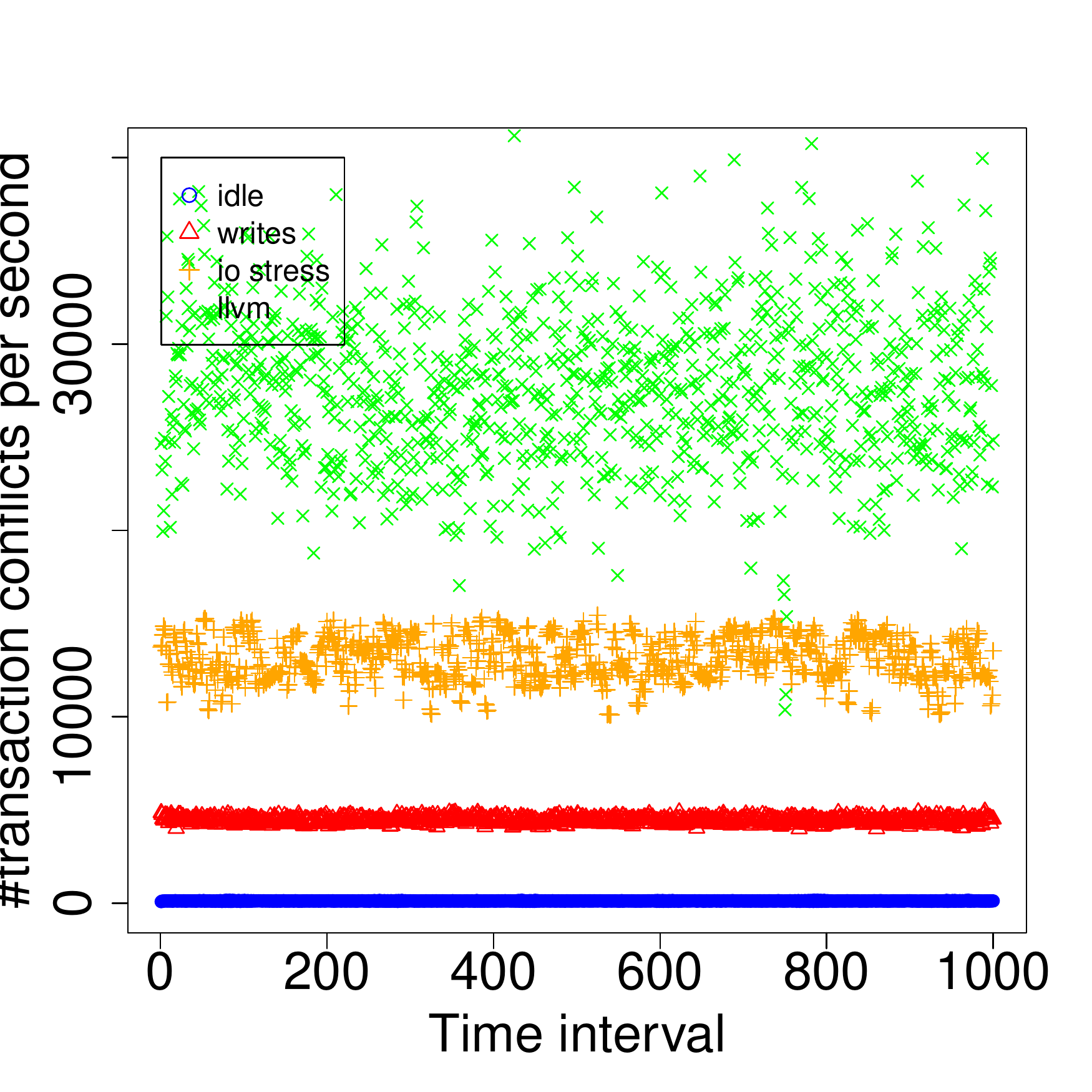}
        \caption{Reading 4~KiB in a transaction significantly increases the chances of memory
        conflicts when the system comes under heavy load, even when executing on a reserved core.}
        \label{fig:rtmAborts:conflicts:read4K}
    \end{subfigure}

    \caption{The number of TSX transaction aborts encountered depends heavily on system load. The 
    graphs display the number of 
    interrupts (Fig.~\ref{fig:rtmAborts:interrupts:whileTrue} and \ref{fig:rtmAborts:interrupts:read4K}) 
    and memory 
    conflicts (Fig.~\ref{fig:rtmAborts:conflicts:whileTrue} and \ref{fig:rtmAborts:conflicts:read4K}) 
    per second under the given 
    system load. All tests were performed on an otherwise reserved core. IRQs were also handled by
    other cores.}
    \label{fig:rtmAbortCounts}
\end{figure*}

\paragraph{RTM Transaction Size}
\label{app:evaluation:tsx:reexecution}
Transactions are constrained both in space and time. Transactions are aborted
when their read or write set grow too large to be tracked or when an interrupt
occurs.
To assess their maximum practical size, we wrote small test programs. 
One version allocates a small buffer and transactionally fills it. When this
transaction succeeds, we immediately attempt the same operation again and record
the chances of this second transaction succeeding immediately. Another version
implements the same test, but waits 1 second between transactions.


Figure~\ref{fig:rtmSuccesses:write} displays the results. 
When the second transactions are started immediately
and the write buffer is smaller than 28~KiB, the success rate of second 
transaction is higher than 90\%. After the 30.5~KiB limit is reached, all transactions fail. 
When we introduce a cool down period of 1 second between transactions, 
success rates remain close to 100\% for buffers smaller than 23~KiB, but
drop sharply afterwards. 

We performed the same test for read instructions. Transactions' read set is tracked
in L1 as well, but cache line evictions ``may not always result in an immediate 
transactional abort since these cache lines may be tracked in an 
implementation-specific second level structure.''~\cite[\S 12.2.4.2]{intel64optimization} 
Our results (see Figure~\ref{fig:rtmSuccesses:read}) show indeed such behavior with
transactions quickly surpassing the L1 cache size. When the buffer grows larger than
2,080~KiB, we again see a big impact of the cool down period. 

We suspect that the impact of the cool down period originates from other processes occupying the same 
cache lines. Processes on other physical cores compete for the (inclusive) L3 cache. 
Processes running on the other logical core of a HyperThread-enabled processor,
compete for the L2 cache as well. We suspect that when our benchmark tries to access a cache line 
that would cause a cache line to be evicted from another logical or physical core in L3, the 
entire transaction is aborted. 



%

\paragraph{Transaction Aborts due to Interrupts}
We wrote a small benchmark to assess the impact of system load on TSX transactions.
Our benchmark starts a new transaction and immediately enters a {\tt while(1)}-loop.
We execute the benchmark on a reserved (physical) core and offload the IRQ interrupts\footnote{
We validated our setup by executing {\tt watch -n1 -d "cat /proc/interrupts"} and watching
interrupts being registered on each core.}
to another core. For a 1,000 second period, we track how many aborts are encountered 
per second and their cause. Unfortunately Intel TSX does only signals a
transaction abort when an interrupt arrives. Therefore we assume all aborts without an
explicitly reported cause, are due to interrupts.
Next, we filter out all non-interrupt related transaction aborts and 
display the result in Figure~\ref{fig:rtmAborts:interrupts:whileTrue}. 
When the system is running idle, 980 transaction abort per second (median).
When the system comes under heavy load, this number increases significantly.
When we instruct the other remaining logical cores to indefinitely write to a 2~MiB 
buffer, 12,854 transactions were interrupted per second.
Performing disk intensive loads using the {\tt stress} tool on the remaining available 
logical cores increases the transaction interrupts per second to 20,282. 
Finally we tested the system under a more realistic load and build the llvm 3.9 compiler on the 
remaining logical cores. This led to a significant increase of 79,067 transaction interrupts per 
second due to interrupts.
When we slightly modify our benchmark to first read a 4~KiB buffer before entering the endless loop,
we received almost the exact same results. This strengthens our assumption that the transactions
were indeed aborted due to interrupts.

\paragraph{Transaction Aborts due to Cache Conflicts}
Using the same results of the previous benchmarks, we also look into the number of aborts 
solely caused by cache conflicts.
Results are displayed in 
Figures~\ref{fig:rtmAborts:conflicts:whileTrue} and \ref{fig:rtmAborts:conflicts:read4K}.
Note the change of scale of the vertical axes. As expected, less than 44 (median) 
transaction aborts per second were caused by cache conflicts when we run the first benchmark 
on an idle system,
or under I/O stress. When the other cores write to a 2~MiB buffer, 
the number of aborts due to cache conflicts increase to 67. While this load is memory
intensive, only a small part of the cache is used. In addition the benchmark only 
uses a very small amount of stack memory to call the {\tt rtm\_begin} function 
(\listing{lst:heisenbergInterface} line \ref{lst:heisenbergInterface@rtmbegin}), 
which explains the modest amount of cache 
conflicts. Compiling the LLVM compiler on the other hand uses many more cache lines, which
results in many more transaction aborts per second (27,913 aborts/s).
As expected, reading a 4~KiB buffer before entering the endless loop, increases the likelihood
of memory conflicts.

\paragraph{Conclusion}
Our benchmarks show that when the system comes under heavy load, chances of transaction aborts 
increase significantly, even when executing on an isolated (physical) core and 
interrupts are offloaded to another core. This has significant impact on {\em all} defense mechanism 
relying on Intel TSX to hook enclave {\tt enclu[eresume]} instructions.

First, it shows that forcing an operating system or hypervisor to guarantee that a certain 
transaction abort limit is not exceeded, is extremely difficult. The entire system load
needs to be taken into account, not just load running on a specific core. This has a
detrimental effect on reactive defense mechanisms that attempt to detect ongoing attacks
based on an unusual amount of transaction aborts.


Second, as transaction aborts depend heavily on system load, wrapping sensitive 
operations in a transaction is challenging. Splitting the operation in small transactions
will lead to significant performance overhead. But too large operations may never terminate
when the system is under heavy load. This makes any solution relying on Intel TSX to hook
enclave resumes, only a short-term solution.

\end{document}